\documentclass[reprint,
superscriptaddress,
nofootinbib,
amsmath,amssymb,
aps,
]{revtex4-1}

\usepackage{tikz}
\usepackage[utf8]{inputenc}
\usepackage{amsmath,amsfonts,amssymb}
\usepackage{graphicx}
\usepackage[toc,page]{appendix}
\usepackage{hyperref}
\usepackage{cleveref}

\usepackage[loose]{units}
\usepackage{booktabs}
\usepackage{float}
\usepackage{mathtools}
\usepackage{multirow}

\definecolor{pyblue}{RGB}{31, 119, 180}

\hypersetup{
    colorlinks=true,
    linkcolor={pyblue},
    citecolor={pyblue},
    urlcolor={pyblue}
}

\makeatletter
\gdef\@fpheader{}
\makeatother

\usepackage{graphicx}
\usepackage{dcolumn}
\usepackage{bm}

\def \vpi{\varphi}
\def\nn {\nonumber}

\def\be{\begin{equation}}
\def\ee{\end{equation}}

\def\ba{\begin{eqnarray}}
\def\ea{\end{eqnarray}}

\def\vpi{\varphi}

\def\beq{\begin{equation}}
\def\eeq{\end{equation}}
\def \cF{{\cal F}}
\def \muma{\mu_\text{max}}
\def \mumi{\mu_\text{min}}

\begin{document}
	
	\preprint{}
	
	\title{Bispectrum Islands}
	
	\author{Claudia de Rham}
    \email{c.de-rham@imperial.ac.uk}
	\affiliation{Abdus Salam Centre for Theoretical Physics, Imperial College, London, SW7 2AZ, UK}
    \affiliation{CERCA, Department of Physics, Case Western Reserve University, 10900 Euclid Ave, Cleveland,
OH 44106, USA}

	\author{Sadra Jazayeri}
	 \email{s.jazayeri@imperial.ac.uk}
	\affiliation{Abdus Salam Centre for Theoretical Physics, Imperial College, London, SW7 2AZ, UK}
    
	\author{Andrew J. Tolley}
    \email{a.tolley@imperial.ac.uk}
	\affiliation{Abdus Salam Centre for Theoretical Physics, Imperial College, London, SW7 2AZ, UK}
    \affiliation{CERCA, Department of Physics, Case Western Reserve University, 10900 Euclid Ave, Cleveland,
OH 44106, USA}
	\begin{abstract}
Inspired by the amplitude bootstrap program, the spirit of this work is to constrain the space of consistent inflationary correlation functions—specifically, the bispectrum of curvature perturbations—using fundamental principles such as unitarity, locality, analyticity, and symmetries. To this end, we assume a setup for inflation in which de Sitter isometries are only mildly broken by the slow roll of the inflaton field, and study the bispectrum imprinted by a generic hidden sector during inflation. Assuming that the hidden sector’s contributions to primordial non-Gaussianity are dominated by the exchange of a scalar operator (which does not preclude high-spin UV completions), we derive nontrivial positivity constraints on the resulting bispectrum $B(k_1,k_2,k_3)$. In particular, we show that $B$ must be negative in a certain region around the equilateral configuration. For instance, for isosceles triangles (with $k_2=k_3$) this region is given by $0.027\lesssim k_3/k_1\leq 2$. Furthermore, we demonstrate that unitarity imposes upper and lower bounds on the bispectrum shape, thereby carving out a \textit{Bispectrum Island} where consistent shapes in our setup can reside. We complement our analysis by contemplating alternative setups where the coupling to the hidden sector is allowed to strongly break de Sitter boosts. We also identify situations that would push the bispectrum off the island and the profound physical features they would reveal.
\end{abstract}
	
	\maketitle

\section{Introduction}

Current observational data on the primordial power spectrum of density perturbations are consistent with the simplest single-field inflationary models \cite{Planck:2018jri}. In these scenarios, a canonical scalar field with a nearly flat potential drives the universe’s exponential expansion during inflation, while its vacuum fluctuations generate an almost scale-invariant spectrum of adiabatic perturbations on super-Hubble scales. Within this framework, the scalar and tensor two-point functions serve as the main observables, providing direct insight into the inflaton potential characteristics. Meanwhile, non-Gaussianities are typically suppressed by slow-roll parameters (see Ref.\cite{Maldacena:2002vr}), making their detection extremely challenging.\footnote{On the optimistic side, see \cite{Munoz:2015eqa} for the tantalizing potential of future 21cm cosmology in achieving percent-level constraints on $f_{\text{NL}}$.}

From a UV perspective, embedding inflation within a fundamental framework such as string theory typically requires the inclusion of additional degrees of freedom—such as Kaluza-Klein excitations, moduli, and other heavy fields \cite{Baumann:2014nda}. These extra states, which we collectively refer to as the \textit{hidden sector}, are generically coupled to the inflaton and can influence the correlation functions of curvature perturbations through exchange processes. Such interactions may be strong enough to produce observable levels of primordial non-Gaussianity, for instance
$f_{\text{NL}}^{\text{eq}}\sim {\cal O}(10)$, within the sensitivity range of current and upcoming observational surveys, including EUCLID, DESI, CMB-S4, LSST, SPHEREx and Simons Observatory \cite{Achucarro:2022qrl}. In this work, we focus on the bispectrum as one of the most promising observables for probing signatures of UV physics during inflation (see e.g. \cite{Alishahiha:2004eh, Tolley:2009fg, Baumann:2011su, Achucarro:2010da, Flauger:2016idt,Pajer:2024ckd}).

If the hidden sector during inflation is gapped, one can formulate a generic effective field theory (EFT) for the massless scalar perturbations, integrating out all heavier degrees of freedom. In this EFT framework, and at leading order in the couplings, the bispectrum is generated by irrelevant cubic operators \cite{Cheung:2007st,Weinberg:2008hq}. However, as we consider in this work, if the hidden sector lies not far above the Hubble scale, its non-local effects can remain imprinted in the bispectrum through exchange processes. When the hidden sector consists of isolated, heavy resonances that are weakly coupled, the corresponding exchange processes lead to distinctive oscillatory signals in the squeezed limit of the bispectrum \cite{Arkani-Hamed:2015bza,Lee:2016vti,Chen:2016uwp,Chen:2016hrz, Wang:2019gbi, Kumar:2019ebj,Bodas:2020yho, Tong:2021wai, Pinol:2021aun,Tong:2022cdz}, which cannot be mimicked by local single-field operators.
More general hidden sectors—such as those that are strongly coupled—cannot be described even by conventional multi-field EFTs, as they may exhibit a continuous spectrum (see, e.g. \cite{Hubisz:2024xnj,Aoki:2023tjm}).

This work aims to chart the landscape of all possible bispectra induced by interactions with a hidden sector during inflation, consisting of degrees of freedom beyond the inflaton and graviton. Our approach is model-independent and, much like the S-matrix positivity bound/bootstrap programs \cite{deRham:2017avq, deRham:2017zjm, deRham:2018qqo,Paulos:2017fhb, Arkani-Hamed:2020blm, Tolley:2020gtv, Caron-Huot:2020cmc,Bellazzini:2020cot, Caron-Huot:2021rmr, Guerrieri:2021tak, deRham:2022hpx,Kruczenski:2022lot}, it is grounded in fundamental principles such as unitarity, locality, analyticity and symmetry.\footnote{See Refs.~\cite{deRham:2017aoj,Grall:2020tqc,Grall:2021xxm,Grall:2022pad,Stefanyszyn:2023qov} for constraints on cosmological EFTs based on unitarity, \cite{Grall:2019qof,Grall:2020ibl,AguiSalcedo:2023nds} for symmetry ones and \cite{Melville:2023kgd,Melville:2024ove,Cespedes:2025dnq} for progress in relating cosmological correlators with their flat counterparts.} These principles are used to directly constrain the observable—the bispectrum—without making explicit reference to the hidden sector’s Lagrangian.

Specifically, we focus on a regime in which the inflaton field $\phi$ is decoupled from metric perturbations and can be treated as a nearly shift-symmetric scalar evolving in a fixed de Sitter (dS) background. In addition, we model the hidden sector with a generic scalar operator ${\cal O}$ that is weakly coupled to $\phi$. With the exception of a few assumptions introduced below, we treat this operator as completely general, without imposing any prior on its microscopic origin. We emphasis that focusing on a scalar operator does not imply that the hidden sector nor the UV completion is limited to a scalar one and in practice our framework encompasses a series of different generic scenarios, including some that involve higher spins states. 

For instance, ${\cal O}$ could correspond to a composite scalar built from other generic fields, a tower of scalar resonances arising from a hypothetical UV completion of inflation, or a strongly coupled sector with a continuous spectrum—such as a conformal field theory defined in the bulk of de Sitter space (see \cite{Green:2013rd, Pimentel:2025rds} for previous work exploring the same possibility). All that is required is that ${\cal O}$ mediates interactions between inflaton fluctuations $\delta\phi$ and contributes to their three-point function. Assuming that this contribution is dominated by the two-point correlator of the hidden sector $\langle {\cal O}{\cal O} \rangle$, we derive constraints on both the sign and shape of the resulting bispectrum. These constraints follow from the positivity of the spectral decomposition of the ${\cal O}$'s two-point function, whose form is fully fixed by unitarity and de Sitter isometries. We also assume that the hidden sector two-point function involves only de Sitter principal series.

While in this work we do not claim to derive the most general bounds on the bispectrum, our goal is to demonstrate that, by leveraging the consequences of unitarity and isometries in rigid de Sitter space, it is in principle possible to impose nontrivial constraints directly on inflationary correlators—much like how S-matrices are constrained in flat space using the same principles in Minkowski space. See also \cite{Hogervorst:2021uvp,DiPietro:2021sjt,DiPietro:2023inn,Penedones:2023uqc,Loparco:2023rug,Loparco:2025azm}, where the same principles are employed to constrain quantum field theories in de Sitter space.

The rest of the paper is organized as follows: In Section \ref{genframe}, we introduce our framework for inflation and state the technical assumptions underlying our bounds. In Section \ref{bisexch}, we derive analytical expressions for the bispectrum arising from the exchange of the operator ${\cal O}$. These results are then used in Sections \ref{bispecsignsize} and \ref{bispecshape} to constrain the sign and shape of the bispectrum. In Section \ref{nontrivial}, we verify that our bounds are satisfied by examining two examples of hidden sectors: one weakly coupled with a finite number of fields, and the other strongly coupled, corresponding to a conformal field theory in the bulk of de Sitter space.
While our bounds rely on specific assumptions, violating them will have profound implications which are discussed in  Section \ref{violation}.  We briefly discuss connections with existing non-Gaussianity templates in Section \ref{Sec:Templates} and possibilities to identify new sets of templates that could potentially capture the current bounds more efficiently. 
More generically, the prospects for deriving further positivity bounds on the inflationary bispectrum and trispectrum are presented in Section \ref{futuredirection}. Finally, in Appendix \ref{appendixA}, we derive positivity bounds for an alternative setup where the coupling to the hidden sector is allowed to strongly break de Sitter boosts—unlike the scenario considered in the bulk of the paper. 

\textit{Notation:} Up to small slow-roll corrections, the spacetime during inflation can be approximated with the Poincaré patch of dS space, charted with the following coordinates
\begin{align}
    \mathrm{d} s^2=\dfrac{1}{\eta^2\,H^2}(-\mathrm{d}\eta^2+\mathrm{d}\bm{x}^2)\,,
\end{align}
where $\eta$ is the conformal time.
Curvature perturbations will be represented by $\zeta$. $\bm{k}$ denotes a comoving 3-momentum, while $k$ stands for its norm. A prime on correlators, e.g. $\langle\zeta_{\bm{k}_1}\dots\rangle'$, indicates that a factor of $(2\pi)^3\delta^3(\bm{k}_1+\dots)$ is stripped off. The power spectrum of $\zeta$ is defined by
\begin{align}
\nn
    P(k)=\langle \zeta_{\bm{k}}\zeta_{-\bm{k}}\rangle'=2\pi^2\,\Delta_\zeta^2/k^3\,,
\end{align}
where $\Delta_\zeta$ denotes the observed amplitude of scalar perturbations ($\approx 2.2\times 10^{-9}$). 
\section{General framework}
\label{genframe}
Let us elaborate on the general framework underlying our bounds. We assume that, to leading order in the slow-roll parameters, the inflaton field can be modeled as an (approximately) shift-symmetric scalar propagating in a fixed de Sitter background, where metric perturbations are ignored (an appropriate decoupling limit is taken as in Eq.~\ref{eq:decoupling}). At the same time, the background expectation value of the inflaton weakly breaks de Sitter boosts,
\begin{align}
    \langle \phi\rangle\approx \dot{\bar{\phi}}\,t\,,
\end{align}
while scale invariance remains preserved due to a specific linearly realized combination of the original shift symmetry of $\phi$ and de Sitter dilatation. In this framework, from the perspective of a low-energy effective field theory, the inflaton self-interactions can be organized as a series of shift-symmetric operators involving one or more derivatives of $\phi$ \cite{Creminelli:2003iq}. The lowest order term in this expansion, in both $\phi$ and its derivatives, is the dimension-8 operator:
\begin{align}
\label{EFTfirst}
    {\cal L}_8=\sqrt{-g}\,\dfrac{c}{\bar{\Lambda}^4}(\partial \phi)^4\,.
\end{align}
To have a consistent EFT expansion for the background trajectory of the inflaton field, it is necessary to assume that $\bar{\Lambda}$ is larger than the symmetry breaking scale $\dot{\bar{\phi}}^{1/2}$, i.e. 
\begin{align}
    \bar{\Lambda}\gtrsim \dot{\bar{\phi}}^{1/2}\,.
\end{align}
The symmetry breaking scale $\dot{\bar{\phi}}^{1/2}$ can be inferred from the observed amplitude of the scalar power spectrum, namely 
\begin{align}
    \Delta_\zeta^2 = \frac{H^4}{4\pi^2 \dot{\bar{\phi}}^2} \sim 2.2 \times 10^{-9}\Rightarrow \dot{\bar{\phi}}^{1/2}\sim 58 H\,,
\end{align}
implying that $\bar{\Lambda}$ is much larger than the Hubble scale. This means that the EFT description of $\phi$, organized as a power series in $(\lambda\bar{\Lambda})^{-1}$, where $\lambda$ is the typical wavelength, remains valid deep inside the horizon (i.e. $58^{-1}H^{-1}\lesssim\lambda\ll H^{-1}$), where a flat-space limit can be defined. In this regime, the EFT of inflation reduces to that of a shift-symmetric scalar in Minkowski space, enabling us to confidently apply the standard flat-space positivity bounds—specifically, that the $\mathcal{O}(1)$ coefficient $c$ must be \textit{positive} \cite{Pham:1985cr,Adams:2006sv}.

De Sitter isometries forbid a non-vanishing three-point function for a shift-symmetric scalar. However, one can weakly break de Sitter boosts by evaluating one external field in the quartic interaction on the background. This procedure generates a dimension-6 cubic interaction for $\delta\phi$, given by 
\begin{align}
    {\cal L}_6=-\sqrt{-g}\,\left(\dfrac{4\dot{\bar{\phi}}}{\bar{\Lambda}^4}\right)c\,\,\delta \dot{\phi}(\partial_\mu\delta\phi)^2\,,
\end{align}
which induces the leading-order three-point function, or bispectrum, of the curvature perturbation ($\zeta=-\frac{H}{\dot{\bar{\phi}}}\delta\phi$) given by  (see Ref.\cite{Creminelli:2003iq})
\begin{align}
\label{bispectrumEFT}
    B(k_1,k_2,k_3)&=\left(\dfrac{c\,H^8}{\bar{\Lambda}^4\, \dot{\bar{\phi}}^2}\right)\dfrac{1}{e_3^3}\dfrac{1}{e_1^2} \\
\nn
    &\times\left(e_1^5-3 e_1^3 e_2-4e_1 e_2^2+11 e_1^2 e_3-4 e_2 e_3\right)\,.
\end{align}
Here, 
\begin{align}
    B(k_1,k_2,k_3)=\langle \zeta(\bm{k}_1)\zeta(\bm{k}_2)\zeta(\bm{k}_3)\rangle'\,,
\end{align}
and $e_{1,2,3}$ are the following symmetric polynomials, i.e. 
\begin{align}
\nn
    & e_1=k_1+k_2+k_3\,,e_2=k_1k_2+k_1 k_3+k_2 k_3\,,\\ 
    & e_3=k_1 k_2 k_3\,.
\end{align}
The shape of non-Gaussianity in this case closely resembles the equilateral template, with the overall amplitude at the equilateral configuration given by
\begin{align}
    f_\text{NL}\equiv \dfrac{5}{18}\dfrac{B(1,1,1)}{(2\pi^2 \Delta_\zeta^2)^2}=-\dfrac{70}{27}\left(\dfrac{c\,\dot{\bar{\phi}}^2}{\bar{\Lambda}^4}\right)\,.
\end{align}
Crucially, the positivity of $c$ implies that $f_{\text{NL}}<0$, while perturbative unitarity requires $|f_{\text{NL}}|\lesssim {\cal O}(1)$. 

This result is significant: assuming de Sitter isometries are only weakly broken by the inflaton vacuum expectation value, the bispectrum shape is fixed at leading order in the EFT expansion, with an equilateral amplitude that must be negative according to flat-space positivity arguments. (In fact, this conclusion holds for all configurations of the bispectrum triangle, i.e. $B(k_1, k_2, k_3) < 0$.)

In this work, we set out to explore how the inclusion of heavy degrees of freedom during inflation modify the bispectrum sign and shape. To that end, we consider a minimal extension of canonical single-field scenarios by introducing a generic hidden sector that is \textit{weakly} coupled to the inflaton. 
This coupling is constructed to respect both de Sitter isometries and the shift symmetry of the inflaton field. 
For simplicity, in this work we restrict our attention to scenarios where the hidden sector couples to the inflaton through a scalar operator ${\cal O}$. We comment on possible generalizations to spinning operators in the final section. 

At leading order in the derivative expansion, the interaction takes the form
\begin{align}
\label{delphi2O}
    {\cal L}_{\text{int}}=-\sqrt{-g}\,\dfrac{\alpha}{\Lambda}(\partial \phi)^2\,{\cal O}\,,
\end{align}
where $\alpha$ is a dimensionless coupling and $\Lambda$ is the cutoff scale. As noted earlier, aside from a few assumptions, we treat ${\cal O}$ as a fully generic scalar operator. The assumptions underlying our bounds are as follows: we require that the correlators of ${\cal O}$ satisfy the Ward identities associated with the de Sitter isometry group $\text{SO}(4,1)$. This is ensured by demanding that the hidden sector is not strongly coupled to the time-dependent background, and can be treated as a spectator operator living in a fixed de Sitter spacetime. 
In particular, we demand that, in the absence of the weak coupling above, the vacuum two-point function of ${\cal O}$ is given by the Källén–Lehmann (KL) spectral decomposition of a scalar operator in $\text{dS}_4$ (see e.g. \cite{Loparco:2023rug}), which takes the form\footnote{Note that this follows from the assumption that the expectation value $\langle {\cal O}(\eta,\bm{x})\rangle$ vanishes. Symmetries require any nonzero $\langle {\cal O}\rangle$ to be constant, so without loss of generality, one can always redefine ${\cal O}$ to make this true.}
\begin{align}
\label{KL}
    \langle {\cal O}(\eta,\bm{x}){\cal O}(\eta',\bm{x}')\rangle=\int_{0}^\infty \mathrm{d}\mu\,\rho(\mu)\,G(\sigma;\mu)\,.
\end{align}
Here, $\mu\equiv\left(\frac{m^2}{H^2}-\frac{9}{4}\right)^{1/2}$ is the mass index, $\sigma$ is the geodesic distance between the bulk points $(\eta,\bm{x})$ and $(\eta',\bm{x}')$, namely
\begin{align}
    \sigma=1-\dfrac{-(\eta-\eta')^2+(\bm{x}-\bm{x'})^2}{2\eta\eta'}\,,
\end{align}
$G(\sigma;\mu)$ is the non-time-ordered two-point function of a free scalar field with mass $m$, given by
\begin{align}
\nn
    G(\sigma;\mu)&=\dfrac{1}{(4\pi)^2}\Gamma\left(\frac{3}{2}+i\mu\right)\Gamma\left(\frac{3}{2}-i\mu\right)\\ &\times\,_2 F_1\left(\frac{3}{2}+i\mu;\frac{3}{2}-i\mu;2,\dfrac{1+\sigma}{2}\right)\,,
\end{align}
and finally $\rho(\mu)$ is the \textit{spectral density}, which must be non-negative in a unitary theory. The positivity of this spectral density drives the bounds that we will obtain in Section \ref{bispecsignsize}. Since the normalization of the operator ${\cal O}$  in \eqref{delphi2O} is degenerate with the coupling $\alpha$, we further assume from now on that the operator ${\cal O}$ is normalized\footnote{For situations in which more subtractions are needed we may choose to normalize $\int_0^\infty \mathrm{d} \mu \rho(\mu) (\mu_0/\mu)^{2n}=1$ for integer $n$ sufficiently large so that the integral converges. For equivalent dimensional arguments it is sufficient to take $\mu_0=H$.} such that $\int_0^\infty \mathrm{d}\mu\,\rho(\mu)=1$. As can be seen from the RHS of the KL decomposition of ${\cal O}$ in \eqref{KL}, we have chosen not to include possible contributions from the complementary series (with masses in the range $(0,3H/2)$). 
In other words, we always assume that $\delta\phi$ is the only light degree of freedom during inflation, while other heavy states (with $\mu>0$) are present in the hidden sector. 

A brief comment on the exclusion of light fields is in order. Since our goal is to constrain scenarios where a single scalar degree of freedom sources the adiabatic perturbations, we must ensure that $\delta\phi$ is the only massless mode in the theory. This requires a gap in the hidden sector spectrum such that $\rho(m) = 0$ for $m < m_\text{gap}$. However, imposing a sharp gap above the threshold $3H/2$ is incompatible with the spectrum of generic interacting QFTs in de Sitter, which at minimum spans the entire principal series. That interactions generically render the spectrum gapless within the principal series is evident already at the perturbative level. A simple example is the one-loop correction to the two-point function of an elementary heavy field in de Sitter, which induces a non-zero spectral density even below the particle production threshold \cite{DiPietro:2021sjt,Hogervorst:2021uvp}. While in principle a sharp gap can be imposed within the complementary series, i.e. $m_\text{gap} < 3H/2$, this comes at the cost of introducing an arbitrary scale $m_\text{gap}$ into the resulting positivity bounds. For this reason, we choose to impose $m_\text{gap} = 3H/2$. This is a natural gap in de Sitter, in the sense that as long as the two-point function of ${\cal O}$ decays faster than $|\sigma|^{3/2}$ at large spatial separations ($\sigma \to -\infty$), the principal series provides a complete basis for the spectral decomposition \cite{Loparco:2023rug}. Furthermore, we expect this gap to be stable under quantum corrections.

Akin to the single-field case reviewed above, we spontaneously break de Sitter isometries by expanding around the (approximately) linear trajectory of the inflaton field, i.e. $\phi=\dot{\bar{\phi}}\,t+\delta\phi$, and find the following interacting terms
\begin{align}
\label{perturbedLag}
    {\cal L}_{\text{int}}=\sqrt{-g}\left(-\dfrac{\alpha}{\Lambda}(\partial_\mu \delta\phi)^2\,{\cal O}+\dfrac{2 \alpha\dot{\bar{\phi}}}{\Lambda}\delta\dot{\phi}\,{\cal O}\right)\,.
\end{align}
The leading observable influenced by the mixing terms above is the $\delta\phi$ two-point function (aka the power spectrum), which receives a correction from the exchange of ${\cal O}$. This correction introduces a small, scale-invariant offset to the zeroth-order power spectrum of the form $A/k^3$, and thus does not encode any distinctive signature of the hidden sector. Therefore, we begin with the three-point function (aka the bispectrum), which reveals significantly more information through its dependence on the shape of the triangle in momentum space.  

The leading-order diagrams contributing to the bispectrum are depicted in Fig.~\ref{fig:O2O3}. They include an exchange process of order ${\cal O}(\dot{\bar{\phi}})$ induced by the two-point function of the operator ${\cal O}$ and another diagram of order ${\cal O}(\dot{\bar{\phi}}^3)$, proportional to the three-point function $\langle{\cal O}^3\rangle$. 
Our next primary ingredient for deriving the bounds on the bispectrum is that the former contribution always dominates. Details about the typical size of each contribution will follow in Section \ref{bispecsignsize}. Our theoretical reason for excluding the three-vertex diagram from our analysis is that, unlike the two-point function of the hidden sector, there is no useful dispersion relation to similarly constrain the analytic structure of the three-point function $\langle \mathcal{O}(x)\mathcal{O}(y)\mathcal{O}(z) \rangle$. To make progress in constraining the final bispectrum, we therefore focus on setups in which the contribution from the three-vertex diagram is subdominant, and the bispectrum can be expressed entirely in terms of the two-point function of $\mathcal{O}$, whose form is fully fixed by the Källén–Lehmann representation.

\section{Bispectrum from Operator Exchange}
\label{bisexch}
\begin{figure}
    \centering
    \includegraphics[scale=0.4]{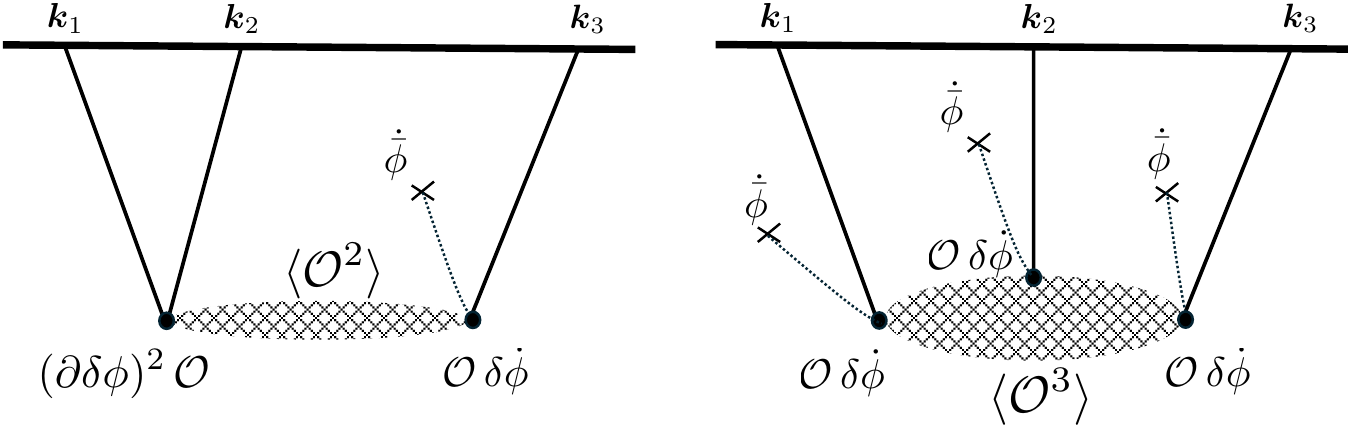}
    \caption{The exchange diagrams contributing to the bispectrum of $\zeta$, induced by the two-point function $\langle {\cal O}{\cal O}\rangle$ (\textit{left}) and  the three-point function $\langle {\cal O}{\cal O}{\cal O}\rangle$ (\textit{right}).} 
    \label{fig:O2O3}
\end{figure}

We employ the in-in formalism to compute the two-vertex exchange diagram shown in Fig.~\ref{fig:O2O3}. The key distinction from standard in-in calculations lies in the treatment of the hidden sector: within the interaction picture, the Wick theorem does not apply to the operator ${\cal O}$. In other words, ${\cal O}$ should not be treated as a free scalar field with a fixed mass, unlike factors of $\delta\phi$, which can be contracted using the usual rules. Instead, at each order in perturbation theory, one must insert the appropriate Green functions for ${\cal O}$. These Green functions are generally unknown unless the full hidden-sector theory is specified, but they are invariant under the action of the de Sitter group.

The diagram of interest involves only two vertices, of either the plus or minus type, so the following are the only Green functions of ${\cal O}$ that enter the computation: 
\begin{align}
\nn
   G^{\cal O}_{++}(\eta,\eta',\bm{x}-\bm{x}')&=\langle 0|\,T\lbrace {\cal O}(\eta,\bm{x}){\cal O}(\eta',\bm{x}')\rbrace|0\rangle\,,\\ \nn
    G^{\cal O}_{--}(\eta,\eta',\bm{x}-\bm{x}')&=\langle 0|\,\bar{T}\lbrace {\cal O}(\eta,\bm{x}){\cal O}(\eta',\bm{x}')\rbrace|0\rangle\,,\\ \nn
    G^{\cal O}_{-+}(\eta,\eta',\bm{x}-\bm{x}')&=\langle 0|\, {\cal O}(\eta,\bm{x}){\cal O}(\eta',\bm{x}')|0\rangle\,,\\
    G^{\cal O}_{+-}(\eta,\eta',\bm{x}-\bm{x}')&=\langle 0|\, {\cal O}(\eta',\bm{x}'){\cal O}(\eta,\bm{x})|0\rangle\,.
\end{align}
These propagators are fully determined by the same Källén–Lehmann representation as in Eq.~\eqref{KL}. In particular, in Fourier space, we find
\begin{align}
   G^{\cal O}_{\pm\pm}(\bm{k},\eta,\eta')=\int_0^{\infty} \mathrm{d}\mu\,\rho(\mu)\,G_{\pm\pm}(\bm{k},\eta,\eta';\mu)\,,
\end{align}
where 
\begin{align}
    G^{\cal O}_{\pm\pm}(\bm{k},\eta,\eta')=\int \mathrm{d}^3\bm{x}\exp(-i\bm{k}\cdot\bm{x})\,G^{\cal O}_{\pm\pm}(\eta,\eta',\bm{x})\,,
\end{align}
are the Fourier transformed bulk-to-bulk propagators associated with ${\cal O}$, and $G_{\pm\pm}(\bm{k},\eta,\eta';\mu)$ are the standard propagators of a fixed-mass scalar in de Sitter: 
\begin{align}
\nn
G_{++}(k, \eta, \eta') &= f_{-}(k, \eta')f_{+}(k, \eta)\theta(\eta - \eta') + (\eta\leftrightarrow\eta'), \\ 
\nn
G_{+-}(k, \eta, \eta') &= f_{+}(k, \eta')f_{-}(k, \eta), \\
\nn
G_{--}(k, \eta, \eta') &= f_{+}(k, \eta')f_{-}(k, \eta)\theta(\eta - \eta') + (\eta\leftrightarrow\eta'), \\
G_{-+}(k, \eta, \eta') &= f_{-}(k, \eta')f_{+}(k, \eta),
\label{props}
\end{align}
with $f_\pm$ standing for the corresponding massive mode functions, namely 
\begin{align}
\nn
f_+(k\eta) &= \frac{\sqrt{\pi} H}{2}e^{-\pi\mu/2 + i\pi/4}(-\eta)^{3/2}H_{i\mu}^{(1)}(-k\eta), \\ \label{sigmaplusminus}
f_-(k\eta) &= \frac{\sqrt{\pi} H}{2}e^{\pi\mu/2 - i\pi/4}(-\eta)^{3/2}H_{i\mu}^{(2)}(-k\eta),
\end{align}
in which $H_{i\mu}^{(1,2)}$ are the Hankel functions. The time integral associated with the exchange diagram depends linearly on $G_{\pm\pm}^{\cal O}$. Therefore, using the KL decomposition of the bulk-to-bulk propagators, it follows that the final bispectrum can also be written as a linear combination of the bispectra induced by the exchange of a set of free heavy states. See also \cite{Marolf:2010zp, Xianyu:2022jwk}, which applies the same spectral method to the computation of one-loop diagrams.

The final spectral decomposition of the bispectrum is given by
\begin{align}
\label{spectralrep}
    B(k_1,k_2,k_3)=\int_0^\infty \mathrm{d}\mu\,\rho(\mu)\,B(k_1,k_2,k_3;\mu)\,,
\end{align}
where $B(k_1,k_2,k_3;\mu)$ is the bispectrum induced by the exchange of a heavy scalar $f$ with the mass index $\mu$, which is coupled to $\delta\phi$ in the same way as the original scalar operator ${\cal O}$ (given in Eq.~\eqref{perturbedLag}). In conclusion, with a known spectral density $\rho(\mu)$, the above representation reduces the computation of the full bispectrum to evaluating the single-particle exchange contributions $B(k_i;\mu)$. 

In recent years, powerful techniques from the \textit{cosmological bootstrap} have yielded useful analytical expressions for $B(k_1, k_2, k_3; \mu)$ \cite{Baumann:2022jpr,Arkani-Hamed:2018kmz}. In particular, it has been shown that the bispectrum exchange diagram can be easily mapped to the exchange diagram of a four-point function involving an auxiliary, conformally coupled field $\vpi$, which interacts with the heavy scalar field $f$ through a cubic term of the form $\vpi^2\,f$. 
Here we only quote the final formulas for this \text{seed} four-point function, which exhibits a much simpler analytic structure compared to the original three-point diagram. The bispectrum will then simply follow by acting with an appropriate \textit{weight-shifting} operator \cite{Baumann:2019oyu}.

It is useful to express this seed four-point first as 
\begin{align}
\nn
     \langle &\vpi(\bm{k}_1)\vpi(\bm{k}_2,\eta_0)\vpi(\bm{k}_3,\eta_0)\vpi(\bm{k}_4,\eta_0) \rangle' \\
     &=\frac{\eta_0^4 H^8}{2k_1 k_2 k_3 k_4}\dfrac{1}{s}\,{\cal F}_4(k_i,s)+(t,u)\,\text{channels},
\end{align}
in which $\eta_0$ is the (conformal) time at which the correlator is evaluated, $s=|\bm{k}_1+\bm{k}_2|$ is the s-channel exchanged momentum, and $\cF$ is given by
\begin{align}
    \cF_4(u,v) &= s\sum_{\pm\pm}\frac{(\pm i)(\pm i)}{2} \\ 
    &\times \int_{-\infty}^{0} \frac{\mathrm{d}\eta}{\eta^2} e^{\pm ik_{12}\eta} \int_{-\infty}^{0} \frac{\mathrm{d}\eta'}{\eta'^2} e^{\pm ik_{34}\eta'}G_{\pm\pm}(s, \eta, \eta')\,.\nn
\end{align}
By rescaling the conformal times in the above integrals, one can explicitly see that $\cF_4$ depends on the external kinematics only through the ratios
\begin{align}
    u=\dfrac{s}{k_1+k_2}\,,\qquad v=\dfrac{s}{k_3+k_4}\,.
\end{align}

It can be shown that, once the seed function $\cF_4$ is computed, the bispectrum of $\zeta$ can be straightforwardly extracted by acting with a suitable weight-shifting operator. In more details, the three-point function can be written in terms of $\cF_4$ as
\begin{align}
\nn
B(k_1,k_2,k_3;\mu) &=\kappa\,\hat{{\cal W}}(k_1,k_2,k_3;\partial_u)\,\cF_4(u,v)\Big|_{u=\frac{k_3}{k_1+k_2},v=1}\\
\label{WSo}
&+(t-u)\,\text{channels},
\end{align}
where
\begin{align}
\label{kappaeq}
\kappa=8\pi^3\alpha^2\dfrac{|\dot{\bar{\phi}}|}{\Lambda^2}\Delta_\zeta^{3}>0\,,
\end{align}
and 
\begin{align}
    \hat{{\cal W}}&=\dfrac{-1}{k_1 k_2 k_3 (k_1+k_2)^3}\left(2\partial_u+u\,\partial^2_u\right)\\ \nn
    &+\dfrac{\bm{k}_1\cdot\bm{k}_2}{k_1^3k_2^3k_3^2}\left(1+u\partial_u+\dfrac{k_1 k_2}{k_3^2}u^3\,\left(2\partial_u+u \partial_u^2\right)\right)\,.
\end{align}
Acting with this operator effectively transforms the bulk-to-boundary propagators of the conformally coupled external states into those of massless ones, while simultaneously adjusting the structure of the vertices. Note that a soft limit is implicitly taken in the expression above by sending the external momentum $\bm{k}_4$ to zero. The relationship between $B$ and $\cF_4$ can be derived straightforwardly from the explicit form of the time integrals; for full details, we refer the reader to \cite{Arkani-Hamed:2018kmz, Jazayeri:2022kjy}.

The seed four-point function can be bootstrapped using a boundary differential equation it satisfies, supplemented by its analytic behavior near its partial energy poles \cite{Arkani-Hamed:2018kmz}. Since the focus of this work is on the bispectrum, we only require the explicit expression for $\cF_4$ evaluated at $v=1$, which takes the following compact form (see \cite{Qin:2023ejc}):
\begin{align}
\nn
    \cF_4(u,v=1)&=\dfrac{1}{\mu^2+1/4}\dfrac{u}{1+u} {}_{3}F_{2}\left(\begin{matrix}
        1,1,1\\ \frac{3}{2}+i\mu,\frac{3}{2}-i\mu
    \end{matrix};\dfrac{2u}{1+u}\right)\\
    \label{fullF}
    &+\dfrac{\pi}{2\cosh(\pi\mu)}h(u)\,,
\end{align}
where 
\begin{align}
\nn
    &h(u)=2i\pi\tanh(\pi\mu)\\ 
    &\times \left((\beta_0-1)\alpha_+\,F_-(u)-\alpha_-(\beta_0+1)\hat{F}_+(u)\right)\,,
\end{align}
and 
\begin{align}
\nn
    \hat{F}_\pm(u)&=\left(\dfrac{i u}{2\mu}\right)^{\frac{1}{2}\pm i\mu}\,_{2}F_1\left[\frac{1}{4}\pm\frac{i\mu}{2},\frac{3}{4}\pm\frac{i\mu}{2},1\pm i\mu,u^2\right]\,,\\ \nn
    \alpha_\pm &=-\left(\dfrac{i}{2\mu}\right)^{\frac{1}{2}\pm i\mu}\dfrac{\Gamma(1\pm i\mu)}{\Gamma(\frac{1}{4}\pm i\frac{\mu}{2})\Gamma(\frac{3}{4}\pm i\frac{\mu}{2})}\,,\\ 
    \beta_0 &=\dfrac{1}{i\sinh(\pi\mu)}\,.
\end{align}
The first term in Eq.~\eqref{fullF} is analytic around $\bm{s} = \bm{k}_1 + \bm{k}_2 = 0$ and encodes local contributions to the four-point function of $\vpi$, arising from a tower of EFT operators generated by integrating out the heavy field at tree level. In contrast, the second term features a branch point at $s = 0$, near which it behaves as $u^{\pm i\mu}$, with a coefficient suppressed by the Boltzmann factor $\exp(-\pi\mu)$, in the large-mass limit. This non-analytic structure captures the pair-production contribution to the four-point function and, upon the action of the weight-shifting operator $\hat{{\cal W}}$ on $\cF_4$, gives rise to the characteristic \textit{cosmological collider oscillations} in the squeezed limit of the bispectrum \cite{Arkani-Hamed:2015bza}.

In the very squeezed limit, the bispectrum asymptotically behaves as
\begin{align}
\nn
    &\lim_{k_3\ll k1\sim k_2}B(k_1,k_2,k_3;\mu)=\kappa \dfrac{1}{k_1^6}\left(\dfrac{k_3}{k_1}\right)^{-3/2+i\mu}\\ 
\label{supersqueezed}
    &\,\,\,\times \beta(\mu)(-15+4\mu(-4i+\mu))2^{-7/2-i\mu}+{\rm c.c.},
\end{align}
where 
\begin{align}
\nn
    \beta(\mu)&=-\dfrac{\pi\mu\,2^{1/2-3i\mu}}{2\cosh(\pi\mu)\Gamma(1+i\mu)}\\
    &\times \Gamma(-i\mu)\Gamma(2i\mu)(-i+\sinh(\pi\mu))\,.
\end{align}
As seen from the above expressions, the amplitude of the cosmological collider oscillations is of order unity for $\mu \lesssim 1$, but becomes exponentially suppressed at large masses. Moreover, the non-Gaussian signal (defined as $\frac{B}{P(k_1)P(k_3)}$) exhibits a power-law suppression in the squeezed limit, scaling as $(k_3/k_1)^{3/2}$, which stems from the dilution of the heavy field in the expanding background.

Note that, even in moderately squeezed triangles, it is essential to account for analytic corrections to the formal asymptotic limit given by Eq.~\eqref{supersqueezed}, appearing at higher orders in $k_3/k_1$. To ensure the accuracy of our numerical results in the following section, we therefore consistently use the exact bispectrum expression rather than relying on the soft-limit approximation. 


\section{The bispectrum sign and size}
\label{bispecsignsize}
\begin{figure}
    \centering
\includegraphics[scale=0.8]{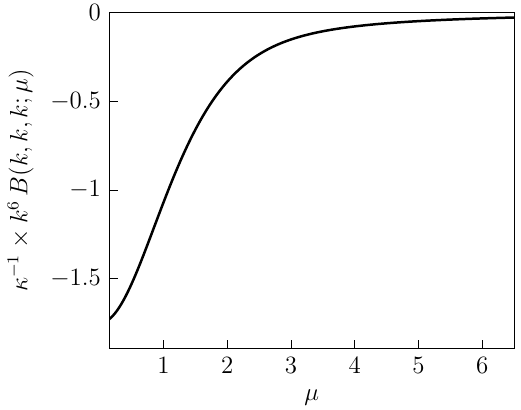}
    \caption{This figure shows that the equilateral bispectrum resulting from the exchange of a heavy field $f$, characterized by mass index $\mu$ and coupled through the perturbed Lagrangian \eqref{perturbedLag} (with ${\cal O}$ replaced by $f$), is negative for all values of $\mu$. Note that the overall coefficient $\kappa$, defined in Eq.~\eqref{kappaeq}, is positive.}
    \label{fig:Bequi}
\end{figure}
The spectral representation in Eq.~\eqref{spectralrep} contains an infinite set of positive coefficients, in principle imposing infinitely many constraints on the bispectrum.
A straightforward example is the sign of the bispectrum in the equilateral configuration ($k_1 = k_2 = k_3$), i.e. 
\begin{align}
    B(k,k,k)=\int_0^\infty \mathrm{d}\mu\,\rho(\mu)\,B(k,k,k;\mu)\,,
\end{align}
which must be \textit{negative}. This property follows directly from the fact that the bispectra associated with the exchange of individual states are negative for all masses (as can be seen in Fig.~\ref{fig:Bequi}), combined with the positivity of the spectral density, $\rho(\mu)>0$. 
As we will soon see, the bispectrum $B(k_1, k_2, k_3)$ 
can become positive for sufficiently squeezed triangles, with its sign oscillating due to the superposition of cosmological collider oscillations (see Eq.~\eqref{supersqueezed}) induced by light fields in the spectrum, i.e. $\mu \lesssim 1$.

Let us make two remarks about the negativity of the bispectrum in the equilateral configuration. First, this sign is insensitive to the overall sign of the coefficient $\alpha$ in front of the original cubic interaction term $(\partial \phi)^2 \mathcal{O}$, as the single-exchange diagram is quadratic in this coupling. Therefore, the initial assumption that the contribution from the three-point function $\langle \mathcal{O}^3 \rangle$ is subdominant plays a central role in judging the sign of $B(k, k, k)$ (see the discussion below). Second, notice that by allowing the hidden sector spectrum to extend to arbitrarily high scales, our analysis automatically captures the inflaton's leading self-interaction $(\partial \phi)^4$, as well as all other subleading terms of the form $(\partial \phi)^2\,\Box^n\,(\partial \phi)^2$ (with $n>1$). These EFT terms emerge from integrating out very heavy states in the spectrum, with coefficients that are necessarily positive.

While we have established the sign of the bispectrum, it is important to emphasize that its overall amplitude is not arbitrarily large in our setup. In particular, the size of the bispectrum is constrained by our assumption that both the cubic and linear mixings in Eq.~\eqref{perturbedLag} remain small at the Hubble scale. This can be ensured by imposing a lower bound on the scale $\Lambda$. The most stringent constraint on this scale arises from requiring the perturbativity of the operator $\delta\dot{\phi}\,{\cal O}$. 
This condition is essential for the consistency of our calculations, which relies on the two-point function of ${\cal O}$ and the mode function of $\delta\phi$ not being significantly modified by this mixing. Note that since this coupling explicitly breaks de Sitter boosts, keeping it small is especially important for preserving de Sitter invariance of the hidden sector two-point function—a key ingredient in deriving positivity bounds in this work. By dimensional analysis, one can ensure that this mixing remains small if $\Lambda \gtrsim \frac{\alpha\dot{\bar{\phi}}}{H}$. As a result, the overall amplitude of non-Gaussianity can be at most of order unity:
\begin{align}
\label{fnlO2}
   |f^{{\cal O}^2}_\text{NL}| \sim {\cal O}(1)\frac{\alpha^2\dot{\bar{\phi}}^2}{\Lambda^2 H^2} \lesssim {\cal O}(1)\,.
\end{align}
The fact that, within our framework, the single-exchange diagram does not generate large non-Gaussianities is not surprising, as so far we did not allow for a sizable breaking of de Sitter boosts. This contrasts with frameworks such as the effective field theory of single- or multi-field inflation \cite{Cheung:2007st,Senatore:2010wk,Noumi:2012vr,Garcia-Saenz:2019njm}, where de Sitter boosts are strongly broken, allowing for $|f^{\text{eq}}_\text{NL}| \gg 1$ to be generated either from $\delta\phi$ self-interactions or from the exchange of other fields (e.g. see \cite{Lee:2016vti}).

In Appendix \ref{appendixA}, we derive positivity bounds on the shape of the bispectrum in scenarios where the coupling to the hidden sector strongly breaks de Sitter invariance, and larger values of non-Gaussianity (i.e. $|f^{{\cal O}^2}_\text{NL}|\gg 1$) can be obtained. 
Specifically, we will consider interactions between the hidden sector and the Goldstone boson of the EFT of inflation $\pi$ of the form
\begin{align}
    {\cal L}_{\text{int}}&=\sqrt{-g}\,c_1(-2\dot{\pi}+(\partial_\mu \pi)^2)\,{\cal O}\\ \nn
    &+\sqrt{-g}\,c_2(-2\dot{\pi}+(\partial_\mu \pi)^2)^2\,{\cal O}\,.
\end{align}
The first building block is effectively captured by our Lagrangian in Eq.~\eqref{perturbedLag}, which originated from the dS invariant block in Eq.~\eqref{delphi2O}, while the second term is typically neglected in a systematic expansion in $(\partial\phi)^2/\bar{\Lambda}^4$. However, in the EFT of inflation, this expansion breaks down because the vacuum is no longer close to the de Sitter invariant one, defined around $\phi = 0$. As a result, the second building block above is not parametrically suppressed relative to the first and must be retained. 
Written in terms of the canonically normalized field\footnote{Here, we are assuming that $\pi$ does not have a non-trivial speed of sound (i.e. $c_s=1$). More generally, the canonically normalized field is not $\delta\phi$ and is instead given by $\pi_c=(2c_s^{-2}M_\mathrm{Pl}^2 |\dot{H}|)^{1/2}\pi$.}  
\begin{align}
\delta\phi=\dot{\bar{\phi}}\,\pi\,, 
\end{align}
this term includes a cubic operator of the form
\begin{align}
\label{newvertex3}
    {\cal L}_{\text{int}}\supset \dfrac{\beta}{\Lambda_*}\delta\dot{\phi}^2\,{\cal O}\,,
\end{align}
the coefficient of which is not tied to the linear mixing term $\delta\dot{\phi}\,\mathcal{O}$. Consequently, the scale $\Lambda_*$—which must exceed the Hubble scale to preserve perturbative unitarity—can, in principle, be much smaller than $\Lambda$. \\

Unlike the single-exchange diagram in the (nearly) de Sitter invariant case, the exchange diagram induced by the new vertex above can lead to large non-Gaussianities of order
\begin{align}
\nn
    \left|f^{{\cal O}^2}_\text{NL}\right|&\sim \dfrac{1}{\Delta_\zeta}\,\left|\beta\dfrac{H}{\Lambda_*}\right|\times \left|\frac{\alpha\dot{\bar{\phi}}}{\Lambda H}\right|\lesssim\dfrac{1}{\Delta_\zeta}\,\,\\
    \label{fnlboost}
    &(\text{strongly broken boosts})\,.
\end{align}
The other difference from the original exchange diagram, which was proportional to $\alpha^2$, is that the final bispectrum is not quadratic in any of the couplings, whose signs are generally unknown. As a result, in the new setup, we cannot constrain the bispectrum sign based solely on the positivity of the ${\cal O}$'s two-point function. In fact, in this case, the contribution from inflaton self-interactions—dominated at leading order by the cubic operators $\dot{\pi}^3$ and $\dot{\pi}(\partial_i \pi)^2$—is also sign-indefinite, in contrast to the bispectrum originating from the dS invariant operator $(\partial \phi)^4$. This difference arises because the usual flat-space positivity bounds do not apply to boost breaking effective field theories (for recent studies in this direction, see \cite{CarrilloGonzalez:2025fqq, Hui:2025aja, Creminelli:2024lhd, CarrilloGonzalez:2023emp, Creminelli:2022onn,Grall:2021xxm,deRham:2021fpu,deRham:2020zyh}). 
To avoid uncertainties from the unknown signs of contact terms, we assume in Appendix~\ref{appendixA} that the contribution from the exchange process dominates. Under this assumption, we will show that the shape of the bispectrum, being independent of the signs of $\alpha$ and $ \beta$, remains subject to constraints derived from the positivity of the two-point function. \\

Finally, we revisit our assumption that the diagram proportional to $\langle {\cal O}^3 \rangle$ in Fig.~\ref{fig:O2O3} is sub-dominant, relative to the single-exchange process. The corresponding amplitude of non-Gaussianity is approximately given by 
\begin{align}
    f_{\text{NL}}^{{\cal O}^3} \sim \frac{1}{\Delta_\zeta} \left( \frac{\alpha\dot{\bar{\phi}}}{\Lambda H} \right)^3 \frac{\langle {\cal O}^3 \rangle_H}{H^3}\,,
\end{align}
where $\langle {\cal O}^3 \rangle_H$ denotes the three-point function of the hidden-sector operator evaluated near Hubble crossing. Compared to $f_{\text{NL}}^{{\cal O}^2}$ (see Eq.~\eqref{fnlO2}), although the above contribution is suppressed by an extra factor of $\dot{\bar{\phi}}/(\Lambda H) \lesssim 1$, it is also enhanced by $1/\Delta_\zeta$. Therefore, to ensure that three-vertex diagram remains sub-dominant compared to the two-vertex one, the three-point function of ${\cal O}$ should be sufficiently small, meaning 
\begin{align}
    \frac{\langle {\cal O}^3 \rangle_H}{\langle{\cal O}^2\rangle_H^{3/2}} \ll \Delta_\zeta \left( \frac{\alpha\dot{\bar{\phi}}}{\Lambda H} \right)^{-1} \sim 10^{-4} \, (f_{\text{NL}}^{{\cal O}^2})^{-1/2}\,.
\end{align}
Here, we have used the fact that the operator ${\cal O}$ is canonically normalized, which implies  $\langle{\cal O}^2\rangle_H\sim H^2$. 

For $f_{\text{NL}}^{{\cal O}^2} \sim \mathcal{O}(1)$, the above condition sets a rather stringent constraint on the operator ${\cal O}$.  
Nevertheless, there are frameworks in which the three-point function of ${\cal O}$ is naturally small, compared to $\langle{O}^2\rangle^{3/2}$. For instance, consider $N$ identical hidden sectors that are not directly coupled, but are interacting with the inflaton sector through an operator of the form $(\partial \phi)^2\,\tilde{{\cal O}}$, where $\tilde{{\cal O}}$ is a canonically normalized operator built out of fields belonging to one copy only. In this case, the effective canonically normalized operator ${\cal O}$, appearing in Eq.~\eqref{delphi2O}, is $\tilde{{\cal O}}/\sqrt{N}$. This implies that
\begin{align}
    \frac{\langle {\cal O}^3 \rangle_H}{\langle{\cal O}^2\rangle_H^{3/2}} \propto 1/\sqrt{N}\,,
\end{align}
showing that the three-vertex diagram can be suppressed relative to the two-vertex diagram when the number of identical sectors interacting with the inflaton field is sufficiently large. This is only but one illustrative proof of principle example which may arise relatively naturally. 

Note that, just like the three-vertex diagram, the single-exchange process induced by the coupling $\delta\dot{\phi}^2\,{\cal O}/\Lambda_*$ is also enhanced by a factor of $\Delta_\zeta$ (see Eq.~\eqref{fnlboost}). As a result, in scenarios where the hidden sector's coupling to the inflaton strongly breaks boost invariance, the single-exchange process can easily dominate—even if $\langle{\cal O}^3\rangle_H$ is of the same order of magnitude as $\langle{\cal O}^2\rangle_H^{3/2}$—as long as
\begin{align}
    \frac{\alpha\dot{\bar{\phi}}}{\Lambda H} \lesssim \left(\frac{H}{\Lambda_*}\right)^{1/2} \lesssim 1\,.
\end{align}
See Appendix \ref{appendixA}, for more details.

\section{Bounds on the bispectrum shape}
\label{bispecshape}
Using the positivity of the bispectrum’s spectral decomposition \eqref{spectralrep}, we can also place constraints on how its amplitude varies as the three-point configuration is deformed. A useful observable that captures this behavior is the shape of the bispectrum, which is conventionally defined by
\begin{align}
    {\cal S}(k_1,k_2,k_3)&=(k_1 k_2 k_3)^2 \dfrac{B(k_1,k_2,k_3)}{B(1,1,1)}\\ 
    &=(k_1 k_2 k_3)^2\dfrac{\int_0^\infty \mathrm{d}\mu\,\rho(\mu)\,B(k_1,k_2,k_3;\mu)}{\int_0^\infty \mathrm{d}\mu\,\rho(\mu)\,B(1,1,1;\mu)}\,. \nn
\end{align}
By this definition, the shape function is normalized in the equilateral configuration, ${\cal S}(k, k, k) = 1$, and is scale-invariant, i.e. ${\cal S}(\lambda k_i) = {\cal S}(k_i)$. As a result, ${\cal S}$ depends on the external kinematics solely through the shape of the corresponding triangle, which can be fully characterized by the two ratios $(k_2/k_1, k_3/k_1)$. 

From the above definition, it is evident that the shape is invariant under an overall rescaling of the spectral density, $\rho(\mu) \to \alpha \rho(\mu)\,\,(\alpha>0)$. In addition, the sign-definite denominator $B(1,1,1)$ is non-zero unless the entire spectral density vanishes. Due to these observations, the bispectrum shape has to be bounded both from above and below at a given configuration $(k_2/k_1,k_3/k_1)$, as the spectral function $\rho(\mu)$ is varied.

To derive the optimal values of ${\cal S}$ for a given configuration $(k_2/k_1,k_3/k_1)$, we first note that $\rho(\mu)$ can always be rescaled so that the denominator is fixed to $-1$. In other words, since the shape is scale invariant, without loss of generality, we can always rescale the density to impose the following constraint on $\rho(\mu)$: 
\begin{align}
\label{linearconstraint}
    \int_0^\infty \mathrm{d}\mu\,\rho(\mu)\,B(1,1,1;\mu)=-1\,.
\end{align}
With this linear constraint, the shape function becomes linear in $\rho(\mu)$, effectively reducing our optimization problem to a linear programming one that can be solved straightforwardly.

For convenience, let us first discretize the spectrum by replacing the spectral integral with a finite sum: 
\begin{align}
   \int \mathrm{d}\mu\,\rho(\mu)\to \sum_{i=1}^{N} \rho_{\mu_i}\,,
\end{align}
where $\mu_i$ are a discrete set of mass indices, and $\rho_{\mu_i}$ are $N$ positive numbers that can be freely varied subject to the constraint $\sum_i \rho_{\mu_i}B(1,1,1;\mu_i)=-1$. 
With this setup, we observe that the intersection of the hyperplane defined by the constraint above and the $N$-dimensional half-spaces, defined by the positivity conditions
$\rho_{\mu_i}\geq 0$, forms a bounded polytope in the $N$-dimensional space where the vector $\vec{v} = (\rho_{\mu_1}, \dots, \rho_{\mu_N})$ lives. According to the fundamental theorem of linear programming, the extrema of any linear function—such as ${\cal S}$—must lie at the vertices of this polytope. In our case, these vertices are located at the corners: 
\begin{align}
    \vec{v}_i=(0,\dots,\rho_{\mu_i},\dots,0)\,,\quad (i=1,\dots,N)\,,
\end{align}
where, by the constraint \eqref{linearconstraint},
\begin{align}
    \rho_{\mu_i}=\dfrac{-1}{B(1,1,1;\mu_i)}>0\,.
\end{align}
In other words, the maximum and minimum values of ${\cal S}$ at a given configuration are determined by the shape of the bispectrum associated with the exchange of a single particle state with a definite mass index $\mu_{\text{min,max}}$. Note that $\mu_{\text{min,max}}$ generally depends on the specific momentum configuration.

\begin{figure}
    \centering
\includegraphics[scale=0.94]{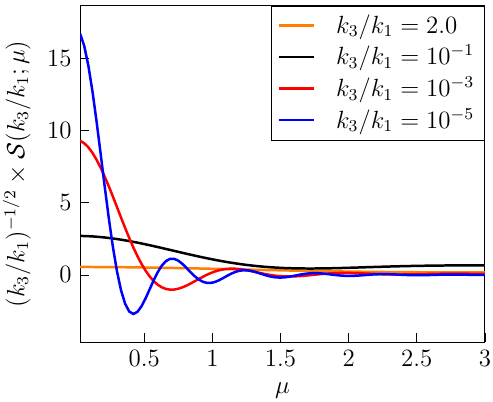}
    \caption{The ansatz for the bispectrum shape ${\cal S}(k_i;\mu)$ is shown as a function of $\mu$, defined by Eq.~\eqref{optimized}. As the plots illustrate across various configurations, the upper bound on the bispectrum shape is always set by the ansatz evaluated at $\mu=0$, while the lower bound ${\cal S}_{\text{min}}$ corresponds to the exchange of fields with masses varying in the range $0<\mu_{\text{min}}<3$.}
    \label{fig:Opmu}
\end{figure}

The  bispectrum shape evaluated at these vertices is given by
\begin{align}
\label{optimized}
{\cal S}(k_2/k_1,k_3/k_1;\mu) 
    \equiv (k_1 k_2 k_3)^2\,\dfrac{B(k_1,k_2,k_3;\mu)}{B(1,1,1;\mu)}\,.
\end{align}
In order to find the optimal values of ${\cal S}$, one has to vary the above ansatz with respect to $\mu\in [0,\infty)$, i.e.  
\begin{align}
\label{optimized2}
\nn
    &{\cal S}_\text{max}(k_2/k_1,k_3/k_1)=\text{max}\left\lbrace
     {\cal S}(k_2/k_1,k_3/k_1;\mu)|\,\mu>0\right\rbrace\\ 
    &{\cal S}_\text{min}(k_2/k_1,k_3/k_1)=\text{min}\left\lbrace
     {\cal S}(k_2/k_1,k_3/k_1;\mu)|\,\mu>0\right\rbrace\,.
\end{align}
Even with this simple final ansatz \eqref{optimized}, the complexity of the bispectrum templates $B(k_i; \mu)$ prevents us from obtaining analytical expressions for the optimal values of the shape function. Instead, for each configuration $(k_2/k_1,k_3/k_1)$, we numerically evaluate the above ansatz and extract the optimal masses $\mu_\text{max}$ and $\mu_\text{min}$, along with the corresponding extrema of the shape function, ${\cal S}_\text{max}$ and ${\cal S}_\text{min}$---see Fig.~\ref{fig:Opmu}.

\begin{figure}
    \centering
\includegraphics[scale=0.92]{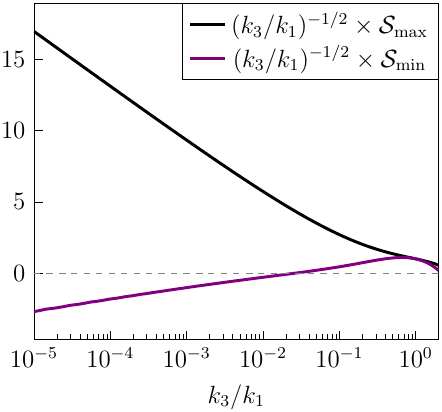}
    \caption{The upper and lower bounds on the shape of the bispectrum ${\cal S}$ plotted as a function of the squeezing parameter $k_3/k_1\leq 2$, for isosceles triangles (with $k_2=k_3$).}
    \label{fig:SmaxSmin}
\end{figure}

\begin{figure}
    \centering
\includegraphics[scale=0.56]{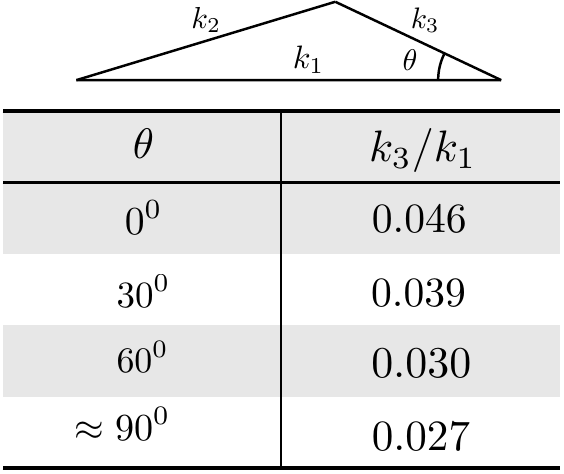}
    \caption{The bispectrum induced by the exchange of the hidden sector must be negative near the equilateral configuration, but it can switch sign for sufficiently squeezed triangles. The table shows the threshold values of the ratio $k_3/k_1$ at which the lower bound on the bispectrum vanishes, as a function of the angle $\theta$ between the short and the long mode.}
    \label{fig:table}
\end{figure}

In Fig.~\ref{fig:SmaxSmin}, we have presented our final bounds ${\cal S}_\text{max,min}$ for isosceles triangles (with $k_2=k_3$), as a function of the squeezing parameter $k_3/k_1$. Let us summarize the general properties of our bounds:   
\begin{itemize}
    \item 
    We have numerically verified that for all triangle configurations the upper bound on ${\cal S}$ is always saturated by the bispectrum shape corresponding to the exchange of a heavy state with mass $m = 3H/2$ (i.e. $\mu = 0$), see Fig.~\ref{fig:Opmu}. In other words, we find $\mu_{\text{max}}=0$ and: 
    \begin{align}
        {\cal S}(k_1,k_2,k_3)<{\cal S}_{\text{max}}={\cal S}(k_1,k_2,k_3;\mu=0)\,.
    \end{align}
    (We also checked that this result remains valid for non-isosceles triangles.)  In particular, in the very squeezed limit ($k_3/k_1\ll 1$), the above maximum bound scales as 
    \begin{align}
    \label{squeezedbehave}
        {\cal S}_{\text{max}}\approx {\cal O}(1)\times (k_3/k_1)^{1/2}\log(k_1/k_3)\,,
    \end{align}
   which matches the squeezed-limit behavior of the bispectrum due to the exchange of a heavy field with $m=3H/2$\footnote{Setting $\mu=0$ in the second part of the seed four-point function (see Eq.~\eqref{fullF}) yields an apparent singularity. However, a careful evaluation of the limit reveals that $\cF$ remains smooth. In particular, one finds a logarithmic behaviour in the squeezed limit of the bispectrum that matches Eq.~\eqref{squeezedbehave}.}. 
    \item In contrast to $\muma$, $\mumi$ depends on the ratio $k_3/k_1$, as can be inferred from Fig.~\ref{fig:Opmu}. In other words, the lower curve in Fig.~\ref{fig:SmaxSmin}, ${\cal S}_{\text{min}}$, does not correspond to the bispectrum shape resulting from the exchange of any mass eigenstate. We also observe that the value of $\mumi$ ranges from 0 (in the squeezed limit) up to $\mu \approx 3.0$, near the folded configuration. 
    \item Moreover, we find that the lower bound on the bispectrum shape remains positive as long as $k_3/k_1 \gtrsim 0.027$, but becomes negative for more squeezed triangles. Combined with the earlier result that the equilateral bispectrum is negative (i.e. $B(1,1,1) < 0$), this implies that the overall sign of the bispectrum is negative for a range of triangle configurations—as long as they are not too squeezed. The corresponding threshold depends on the angle between the short and the long mode, see Fig.~\ref{fig:table}.
\end{itemize}

\section{Examples of explicit scenarios}
\label{nontrivial}
\begin{figure}
    \centering
\includegraphics[scale=1.0]{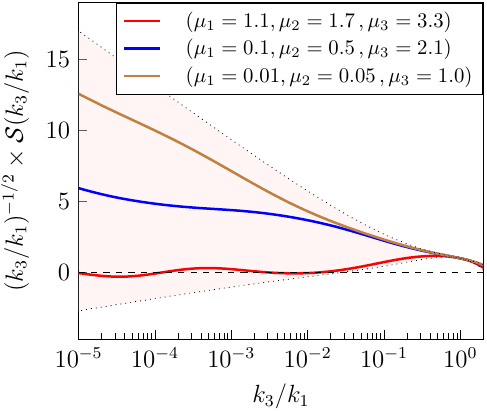}
    \caption{Shapes of the bispectrum generated by the superposition of exchanges of three heavy fields $f_i (i=1,2,3)$ with different masses $\mu_i$. All fields are assumed to couple to the inflaton field with equal strength, via the cubic operator $(\partial \phi)^2\,f_i$.}
    \label{fig:tripleresonance}
\end{figure}

Our bounds must be satisfied by any bispectrum resulting from the tree-level exchange of a unitary sector coupled to the inflaton field as specified in \eqref{delphi2O}. The simplest scenario to test involves a linear superposition of a finite number of resonances in the hidden sector. In Fig.~\ref{fig:tripleresonance}, we illustrate a few examples with three such resonances and confirm that the resulting bispectrum shapes lie within the allowed region. We emphasize again that, as shown by the red curve in the figure, the bispectrum can indeed change sign for sufficiently squeezed triangle configurations. This behavior arises from the cosmological collider oscillations dominating over the local contributions to the bispectrum (see Eq.~\eqref{supersqueezed}).

As another non-trivial example, we take the hidden sector to be a conformal field theory(CFT) defined in the bulk of de Sitter space, with ${\cal O}$ representing a scalar CFT operator of the scaling dimension $\delta$. (Such CFT should not be confused with a hypothetical CFT defined on the future boundary of de Sitter space, which might be dual to the theory in the bulk \cite{Strominger:2001pn}.) 

In this scenario, the two-point function is entirely fixed by conformal invariance and is given by
\begin{align}
    \langle {\cal O}(\eta,\bm{x}){\cal O}(\eta',\bm{x}')\rangle=\dfrac{(-\eta)^\delta(-\eta')^\delta}{\left[(\bm{x}-\bm{x}')^2-(\eta-\eta')^2\right]^\delta}\,.
\end{align}
While unitarity requires the conformal dimension to satisfy $\delta > 1$, it has been shown \cite{Hogervorst:2021uvp,Loparco:2023rug} that for $\delta > 3/2$, the corresponding Källén–Lehmann decomposition can be fully expressed in terms of principal series, with the following spectral density:
\begin{align}
\nn
    &\rho_\delta(\mu)=\dfrac{2^{4-2\delta}\pi}{\Gamma(\delta)\Gamma(\delta-1)}\mu\,\sinh(\pi\mu)\\ 
\label{spectraldensity}
    &\times\Gamma(\delta-3/2-i\mu)\,\Gamma(\delta-3/2+i\mu)\,.
\end{align}
We therefore expect that the bispectra arising from the exchange of such CFT operators should respect our bounds, provided that $\delta > 3/2$. However, a technical subtlety must be addressed before drawing this conclusion. Our derivation relies on the convergence of the spectral integral in the $\mu \to \infty$ limit, which in turn requires the spectral density to grow sufficiently slowly in the UV to ensure convergence. As shown in the next section, when the spectral integral diverges, our bounds no longer apply, and the setup must be re-analyzed using a suitably subtracted spectral decomposition.

To constrain the UV behavior of the spectral density, we first note that in the large-mass limit, the bispectrum induced by a mass eigenstate behaves as
\begin{align}
    \lim_{\mu\to \infty}B(k_1,k_2,k_3;\mu)=B_{\text{EFT}}(k_1,k_2,k_3;\mu)\propto \dfrac{1}{\mu^2}\,,
\end{align}
where $B_{\text{EFT}}$ is the bispectrum generated by the contact interaction 
\begin{align}
    {\cal L}_{\text{EFT}}=-\sqrt{-g}\frac{4 \alpha^2\dot{\bar{\phi}}}{\Lambda^2\,m^2}\delta\dot{\phi}(\partial_\mu\delta\phi)^2\,,
\end{align}
arising from integrating out a heavy scalar with mass $m$. As a result, the spectral integral remains convergent as long as the spectral density is polynomially bounded in the UV such that $\lim_{\mu\to \infty}\rho(\mu)/\mu=0$.

In the case of a CFT operator with conformal dimension $\delta$, the spectral density asymptotically behaves as $\rho(\mu) \propto \mu^{2\delta - 3}$. This implies that, for the convergence of the spectral integral, we must have $\delta < 2$. In conclusion, we expect our bounds to hold for the exchange of scalar operators with dimensions in the range $3/2 < \delta < 2$, where the two-point function can be expressed as a UV-finite integral over the mass spectrum of heavy fields in principal series. 

As an aside, note that when focusing solely on the non-analytic part of the bispectrum in the squeezed limit, as given by Eq.~\eqref{supersqueezed}, the spectral integral converges for any value of $\delta$. This is because, at large masses, the amplitude of the squeezed-limit oscillations scales as $\exp(-\pi\mu)$—in contrast to the $1/\mu^2$ scaling of the analytic part. Consequently, even for $\delta > 2$, the spectral decomposition remains convergent due to the exponential suppression of the UV tail.  
\begin{figure}
    \centering
\includegraphics[scale=0.8]{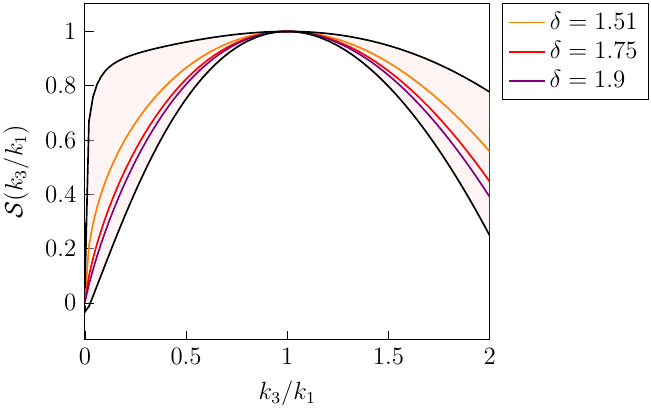}
    \caption{Bispectrum shapes induced by the exchange of scalar CFTs with three sample dimensions within the allowed range $\delta\in (3/2,2)$, where our positivity bounds apply.}
    \label{fig:CFT}
\end{figure}

In Fig.~\ref{fig:CFT}, using the spectral density given in Eq.~\eqref{spectraldensity}, we have numerically evaluated the bispectrum arising from the exchange of a scalar CFT operator for three sample values of the conformal dimension $\delta \in (3/2, 2)$. As shown in the plots, the resulting shapes lie within the allowed region. 

\section{Lost at sea?}
\label{violation}
Let us discuss what conclusions can be drawn about the physics of inflation\footnote{It might be possible to place similar positivity bounds on the bispectrum in alternatives to inflation, for instance bouncing cosmologies \cite{Khoury:2001bz,Khoury:2001wf,Gasperini:2002bn,Steinhardt:2001st,Steinhardt:2002ih,Khoury:2003rt,Tolley:2007nq,Creminelli:2007aq,Cai:2012va,Brandenberger:2012zb,Battefeld:2014uga,Xue:2013bva,Brandenberger:2016vhg,Ijjas:2016vtq,deRham:2017aoj,Tukhashvili:2023itb}, if the precise relationship between $\zeta$ in the expanding phase and the quantum fields (e.g., in the contracting phase) is known, and if the nonlinearities arising through the bounce are under theoretical control.} if future observational experiments are proven to be off the predicted island -- in other words if future observations confirm a primordial non-Gaussianity signal, with an amplitude of order one ($|f^{\text{eq}}_{\text{NL}}| \lesssim 1$), with a sign or shape that violate our bounds.   

Our bounds are based on the general framework outlined at the beginning. In particular, our analysis assumes approximate de Sitter invariance, which provides systematic control over the coupling between the hidden sector and the inflaton fluctuations. We also assumed a weak coupling to the hidden sector, allowing us to compute the bispectrum using the in-in formalism at leading order in perturbation theory. Furthermore, we focused on single-exchange processes (of the scalar type) as the dominant contribution to non-Gaussianity, neglecting diagrams proportional to $\langle {\cal O}^3 \rangle$ and those with the exchange of higher spin operators (see Section \ref{futuredirection}). Deviating from any of these assumptions can lead to a breakdown of our bounds on the sign and/or shape of the bispectrum. Here, we instead highlight a few single-exchange processes in setups with approximate de Sitter invariance that nevertheless violate our bounds, and in each case, we explain the underlying reason.\\

\noindent $\bullet$ We start with the most exotic scenario—namely, coupling to a non-unitary hidden sector. Unitarity is violated, for instance, if some components of the spectral density are negative. In such case, the exchange of the corresponding operator ${\cal O}$ can lead to e.g.  positive contributions to the bispectrum near the equilateral configuration, in contrast to our prediction that $B(1,1,1)$ must be negative. A concrete example is the exchange of a CFT operator with a wrong scaling dimension, i.e. $\delta < 1$, for which $\rho(\mu)<0$ (as can be directly seen from Eq.~\eqref{spectraldensity}). 
In this case, the bounds on the bispectrum shape are still respected since the shape is invariant under $\rho\to -\rho$.   \\

\noindent $\bullet$ A second, less extreme example is a unitary hidden sector with a spectral density that grows too quickly in the UV, causing the spectral decomposition of the bispectrum to diverge. In this scenario, the original spectral decomposition of the bispectrum \eqref{spectralrep} no longer applies, and therefore, neither do our bounds. In order to obtain a useful expression for the bispectrum in the presence of such UV divergences, one must modify the original spectral decomposition of the two-point function by appropriately renormalizing the operator ${\cal O}$. At the level of the exchange diagram for the bispectrum, this procedure is exactly equivalent to absorbing the divergent part of the spectral integral into a series of local counterterms within the effective action of $\delta\phi$. These counterterms take the following form:
\begin{align}
    {\cal L}^{(N)}_{\text{ct}}=\sum_{n=1}^N c_n\,(\partial_\mu\delta\phi)^2\,\Box^{(n-1)}\delta\dot{\phi}\,,
\end{align}
where $N$ is the number of required counterterms, which depends on the degree of the UV divergence of the spectral density $\rho(\mu)$. Once these counterterms are added, the subtracted spectral decomposition of the renormalized bispectrum can be expressed as
\begin{align}
\nn
    B(k_i)&=\sum_{n=1}^N c^R_n\,B_n(k_i)\\ 
\label{subtractedbispectrum}
    &+\int_0^\infty \mathrm{d}\mu\,\rho(\mu)\,\left[B(k_i)-\sum_{n=1}^N d_n(\mu)B_n(k_i)\right]\,,
\end{align}
where $B_n(k_i)$ is the bispectrum associated with the cubic contact term $(\partial_\mu\delta\phi)^2\,\Box^{(n-1)}\delta\dot{\phi}$, which is multiplied by the mass-dependent coefficient
\begin{align}
    d_n(\mu)=-\dfrac{4\alpha^2\dot{\bar{\phi}}}{\bar{\Lambda}^2}\,\dfrac{1}{(\mu^2+9/4)^{n}}\,.
\end{align}
These coefficients are the same as those appearing in the EFT of $\delta\phi$, when a very heavy state (with the mass index $\mu\gg 1$) is integrated out. 
The renormalized coefficients $c_n^R$ above are the sum of two (a priori infinite) quantities 
\begin{align}
    c_n^R=c_n+\sum_{n=1}^N\int_0^\infty \mathrm{d}\mu\,\rho(\mu)d_n(\mu)\,.
\end{align}
Assuming that $\rho(\mu)$ is polynomially bounded, there exists a minimum number of subtractions $N$ such that $\rho(\mu)/\mu^{2N+1} \to 0$ as $\mu \to \infty$. With this choice of $N$, the $\mu$-integral in the second term of Eq.~\eqref{subtractedbispectrum}—which involves positive coefficients—becomes convergent. In contrast, the renormalized coefficients $c_n^R$ are not necessarily positive, as they result from adding two divergent, sign-indefinite quantities. Therefore, the final expression \eqref{subtractedbispectrum} cannot be directly used to constrain the sign or shape of the bispectrum in the same way as in the unsubtracted case. For example, when only one subtraction is required ($N=1$), one could encounter a situation where the first contact term $c^R_1 B_{1}(k_i)$ appears with the wrong sign $c^R_1>0$ and dominates the bispectrum, thereby yielding a positive value for $B(1,1,1)$.  

It would be interesting to explore whether there exists a subtracted observable—constructed from the bispectrum—from which the sign-indefinite terms (i.e. $\sum_{n=1}^Nc^R_n\,B_n(k_i)$) are effectively projected out. One could then impose positivity constraints on this subtracted quantity, rather than on the bispectrum itself. We leave the study of this possibility for future work. \\

\noindent $\bullet$ The third example involves the exchange of light fields (see, e.g., \cite{Chen:2009zp}), which leads to a violation of the upper bound on ${\cal S}$. 
This can be seen from the squeezed limit behaviour of the bispectrum due to the exchange of a single particle state with mass $m<3H/2$, given by
\begin{align}
    S\propto (k_3/k_1)^{1/2-\sqrt{9/4-m^2/H^2}}\,.
\end{align}
Clearly, for sufficiently small $k_3/k_1$, the shape grows larger than $S_\text{max}$. At the same time, we find that the corresponding shapes still remain above our lower bound ${\cal S}_{\text{min}}$. Therefore, adding a set of isolated light states to the hidden sector spectrum simply pushes the upper bound on the shape to a new limit set by the shape associated with the exchange of the lightest state, while the lower limit remains unchanged. 
It is worth noting that in generic multi-field scenarios, de Sitter invariance can be strongly broken by the background trajectory in field space (see, e.g., \cite{Achucarro:2010da,Renaux-Petel:2015mga}); therefore, even our lower bound on the shape is not applicable in such setups. 

Closely related to the previous example is the bispectrum generated by the exchange of a CFT operator with dimension $\delta \in [1,3/2)$. The shape of the bispectrum in this case exhibits the following squeezed-limit behavior (\cite{Green:2013rd}):
\begin{align}
    {\cal S}\propto \left(k_3/k_1\right)^{\delta-1}\,,
\end{align}
which will eventually violate the upper bound ${\cal S}_{\text{max}}$ for sufficiently squeezed triangle. This is of course not surprising, as the spectrum of the CFT operator ${\cal O}$ in this range of dimensions also includes a single light particle state with mass $m^2=\delta (3-\delta)$ \cite{Hogervorst:2021uvp}.

\section{Typical non-Gaussianity templates and beyond}
\label{Sec:Templates}
\begin{figure}
    \centering
\includegraphics[scale=0.72]{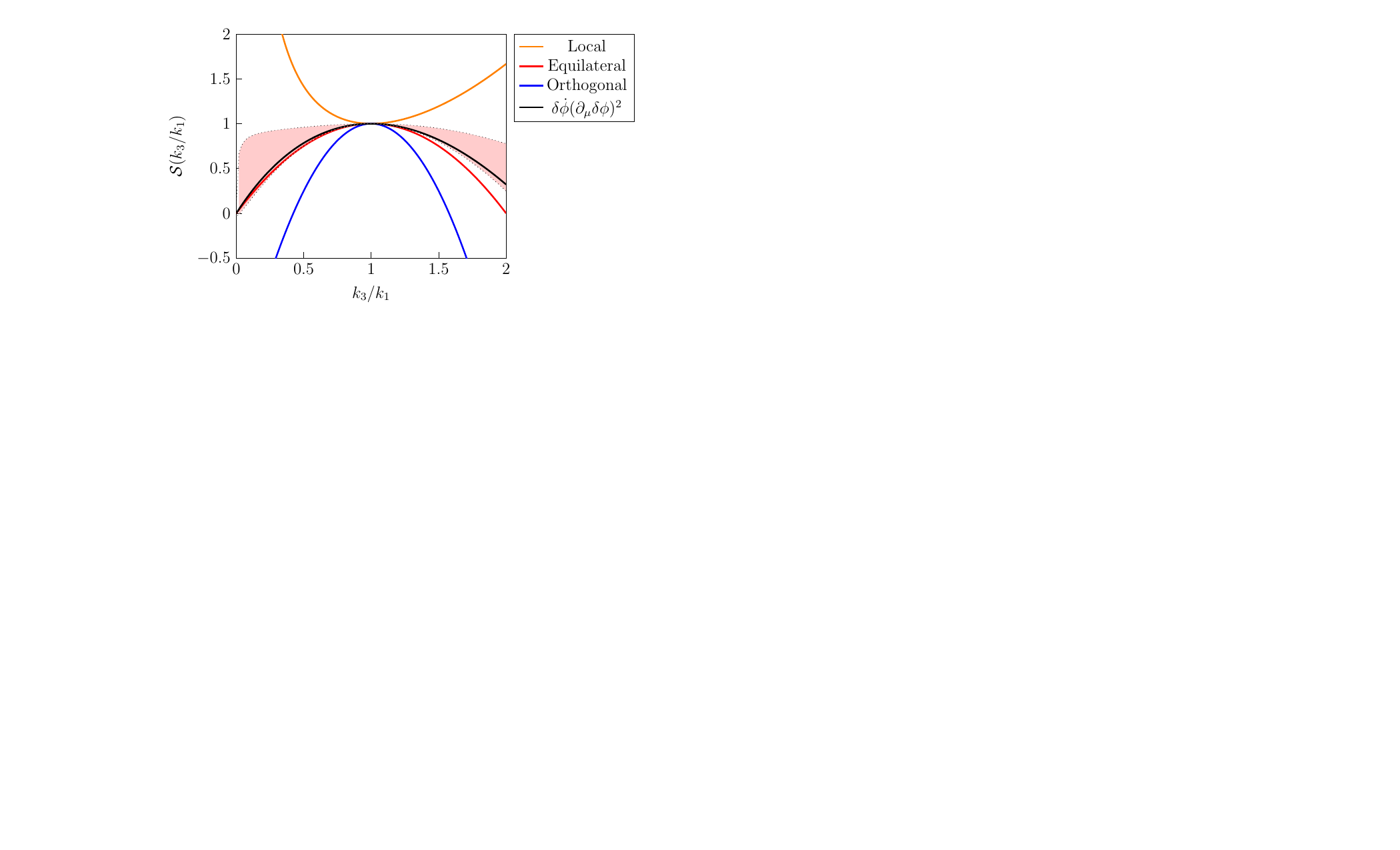}
    \caption{The figure shows that the standard non-Gaussianity templates (Local, Orthogonal, and Equilateral) are inconsistent with our bounds, while the shape generated by the leading-order EFT operator $\delta\dot{\phi}(\partial \delta\phi)^2$ lies within our bispectrum island. See the discussions in Section \ref{Sec:Templates}. Note that, compared to the logarithmic x-axis of Fig.~\ref{fig:SmaxSmin}, the horizontal axis is linear in $k_3/k_1$. The pink area represents the same region between the two curves in Fig.~\ref{fig:SmaxSmin}, which are multiplied by $(k_3/k_1)^{-1/2}$ on the y-axis.}
    \label{fig:Bequi2}
\end{figure}

In this work, rather than predicting specific shapes of non-Gaussianity, we identified a region where 
consistent shapes can reside. Figure \ref{fig:Bequi2} shows this peninsula-like region, restricted to isosceles configurations of the bispectrum triangle. It is worth noting that the conventional local, equilateral, and orthogonal templates used in the CMB bispectrum analyses, entirely or partially lie outside this region. This is of course not very surprising, as the typical scenarios that predict shapes resembling these templates violate our underlying assumptions in one way or another. Specifically, the local template is highly correlated with the shape of non-Gaussianity generated by the exchange of a light field---a scenario we excluded from the outset by imposing a gap at $m=3H/2$. Moreover, the equilateral and orthogonal templates are well captured by shapes generated from appropriate linear combinations of the cubic terms in the EFT of inflation, namely $\delta\dot{\phi}^3$ and $\delta\dot{\phi}(\partial_i \delta\phi)^2$ \cite{Senatore:2009gt}.
Such linear combinations arise only in the presence of strongly broken boosts; therefore, the resulting bispectra are not expected to satisfy our bounds. 
Conversely, as discussed in Section \ref{genframe}, requiring that boosts be only weakly broken selects the unique combination  $\delta\dot{\phi}(\partial_\mu\delta\phi)^2$, whose corresponding shape is shown by the black curve in Fig. \ref{fig:Bequi2}, which falls within the peninsula. 

It should be emphasized that although the equilateral and orthogonal templates themselves are outside the allowed region, they still provide a useful basis for effectively capturing many of the curves within the peninsula (corresponding to different choices of the spectral density $\rho(\mu)$). As a result, existing observational bounds on $f_{\text{NL}}^{\text{eq}}$ and $f_{\text{NL}}^{\text{orth}}$ can already be used to constrain the amplitude of such curves, along the lines of \cite{Cabass:2024wob}. 

Looking ahead, it would be interesting to identify a sufficient set of templates to effectively capture the diversity of possible shapes encompassed by the peninsula (in this direction, see the recent work \cite{Sohn:2024xzd} where a variety of templates associated with the exchange of massive fields are compared with the Planck CMB data---though some of the studied examples lie outside the scope of our assumptions, for instance due to strong breaking of boosts \cite{Pimentel:2022fsc,Jazayeri:2023xcj,Jazayeri:2022kjy}).  
Beyond the standard template-based analyses, it is also important to consider how best to analyze the bispectrum data to assess whether it favors or disfavors the carved region as a whole. We leave this question for future investigations. 
\section{Future Directions}
\label{futuredirection}
In this paper, we presented new perspectives on placing non-trivial, model-independent bounds directly on inflationary correlators, grounded in the general principles of unitarity, analyticity, and symmetry. We focused on a framework in which the starting point is the well-understood dispersion relation for the two-point function of an operator ${\cal O}$, which resides in a hidden sector during inflation and is expressed as a function of the geodesic distance $\sigma$. This dispersion relation is governed by the Källén--Lehmann spectral representation in de Sitter space. Assuming that the bispectrum at leading order is sourced by the exchange of this operator, we derived a useful decomposition of the bispectrum as a sum over contributions from the exchange of an infinite spectrum of single-particle states in de Sitter. Crucially, the coefficients in this expansion are positive due to unitarity, allowing us to derive positivity bounds on the sign and shape of the bispectrum.\\

Here, we highlight a few directions for future study:
\begin{itemize}
    \item 
    While we focused on bounding the value of the bispectrum shape at a fixed momentum configuration, it is equally valuable to study the correlations between shapes arising from different spectral densities. To this end, one can use inner products motivated by observational searches (e.g.,~\cite{Babich:2004gb}) to establish positivity constraints on the overlaps between these shapes and a chosen set of templates. 
    These overlaps involve the bispectrum inner product with itself, which is quadratic in the spectral density and therefore leads to a more challenging non-linear optimization problem. 
    \item The hidden sector we considered may arise from new degrees of freedom within a UV completion of inflation, which typically involves extra dimensions. In such scenarios, one encounters a tower of spin-1 and spin-2 Kaluza-Klein modes coupled to the inflaton. Motivated by this possibility, it would be interesting to extend our analysis to include generic spin-1 and spin-2 operators interacting with the inflaton field (see also \cite{Biagetti:2017viz,Dimastrogiovanni:2018uqy,Kolb:2023dzp} for more generic spin-2 effects). 
    \item Our analysis relied on treating the coupling to the hidden sector perturbatively. This assumption was crucial for identifying the leading-order interaction between the inflaton perturbations and the hidden sector, which ultimately allowed us to compute a three-point function that slightly breaks de Sitter invariance. Formulating a fully non-perturbative bootstrap for the bispectrum that systematically incorporates de Sitter-breaking interactions is not straightforward. However, one can envision scenarios in which de Sitter boost-breaking interactions are subleading, allowing the leading-order \textit{trispectrum} to remain approximately de Sitter invariant. In such cases, it may be possible to develop a non-perturbative bootstrap for the trispectrum of a shift-symmetric scalar in a rigid de Sitter space, leveraging the recently developed partial-wave expansion for the four-point function \cite{Hogervorst:2021uvp,DiPietro:2021sjt}.

\end{itemize}

\textit{Acknowledgments} We would like to thank Giovanni Cabass, Sebastián Céspedes,  Calvin Yi-Ren Chen, Carlo Contaldi, Enrico Pajer, Guilherme L. Pimentel, Arthur Poisson, David Stefanyszyn, S\'ebastien Renaux-Petel, Xi Tong, Kamran Salehi Vaziri, Dong-Gang Wang, Denis Werth, Chen Yang and Yuhang Zhu for stimulating discussions. We also thank Borna Salehian for comments on the manuscript. CdR and AJT are partially supported by the STFC Consolidated Grant ST/X000575/1. CdR and SJ are also supported by Simons Investigator Award 690508.
\begin{figure}
    \centering
    \includegraphics[width=1.02\linewidth]{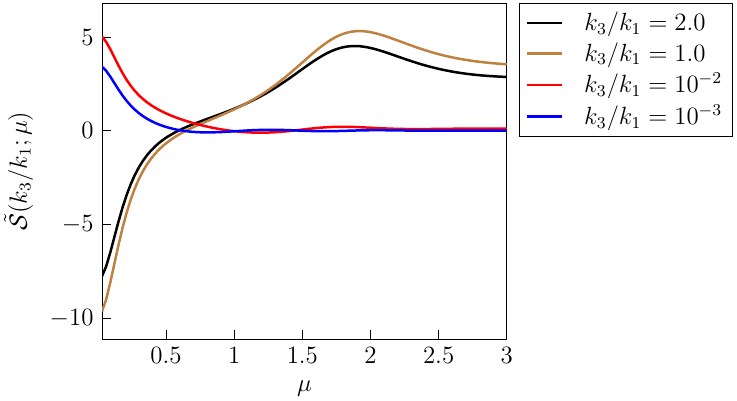}
    \caption{The bispectrum shape ansatz defined in Eq.~\eqref{newansatz}, shown as a function of $\mu$. Unlike in Type I, the maximum value of the bispectrum shape—defined differently here via Eq.~\eqref{newshape}—is not always saturated by the bispectrum induced by the exchange of the lightest particle ($\mu = 0$). Instead, we observe that $\mu_{\text{min}}$ and $\mu_{\text{max}}$ depend on the configuration and vary between 0 and $\approx 3$.}
    \label{fig:Opm2}
\end{figure}
\appendix
\section{Large Boost Breaking Interactions}
\label{appendixA}
\begin{figure}
    \centering
    \includegraphics[width=0.5\linewidth]{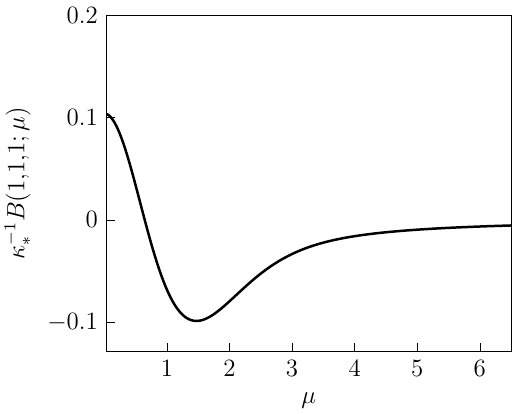}
    \includegraphics[width=0.478\linewidth]{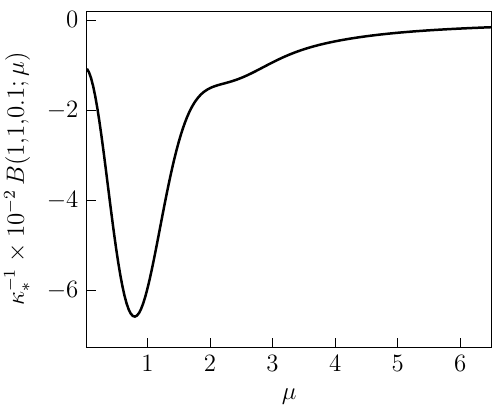}
    \caption{The \textit{left} panel shows that in the Type II setup studied in Appendix \ref{appendixA}, the bispectrum template \( B(k_i; \mu) \) evaluated at the equilateral configuration changes sign as \( \mu \) varies. In contrast, for certain triangle shapes—such as \( (1, 1, 0.1) \)—the sign remains fixed for all values of \( \mu \) (\textit{right}). See Appendix \ref{appendixA} for further discussion.}
    \label{fig:TypeBtemplates}
\end{figure}
In this appendix, we derive positivity bounds on the bispectrum shape in a setup where de Sitter boosts are strongly broken due to coupling with the hidden sector. For clarity, we refer to the original and new setups as Type I and Type II, respectively. Importantly, in the Type II case, we continue to assume that the interaction with the inflaton field remains weak enough for the hidden sector to be treated approximately de Sitter invariant.

To couple scalar perturbations in the Type II setup to the hidden sector, we begin by writing down the leading order interactions
in the unitary gauge, where $\delta\phi=0$. In this gauge, scalar perturbations are absorbed into $\delta g^{\mu\nu}$, and the Lagrangian up to quadratic order in perturbations is given by (see e.g. \cite{Green:2013rd})
\begin{align}
    {\cal L}_{\text{int}
    }=\sqrt{-g}\left( c_1 \delta g^{00}{\cal O}+c_2 (\delta g^{00})^2 {\cal O}+\dots\right)\,,
\end{align}
where dots stand for higher order terms in perturbations or in derivatives (including extrinsic curvature terms, etc). In the decoupling limit, defined by  
\begin{align}
\label{eq:decoupling}
M_\mathrm{Pl}\to \infty\,,\,\,\dot{H}\to 0\, \,\,(M_\mathrm{Pl}^2|\dot{H}|=\text{const.})\,,
\end{align}
the effective action can be expressed in terms of the Goldstone boson $\pi$ by performing the time diffeomorphism $t\to t-\pi(\eta,\bm{x})$, resulting in 
\begin{align}
    {\cal L}_{\text{int}}&=\sqrt{-g}\,c_1(-2\dot{\pi}+(\partial_\mu \pi)^2)\,{\cal O}\\ \nn
    &+\sqrt{-g}\,c_2(-2\dot{\pi}+(\partial_\mu \pi)^2)^2\,{\cal O}\,.
\end{align}
For simplicity, let us assume that the inflaton sector consists of a canonical scalar field, so that $\pi = \delta\phi / \dot{\bar{\phi}}$. The first building block in the Lagrangian above, when expressed in terms of $\delta\phi$, matches the terms already present in Eq.~\eqref{perturbedLag}. The second building block, however, introduces a new cubic vertex,
\begin{align}
\label{Lambdastartverrex}
    \Delta {\cal L}=\sqrt{-g}\,\dfrac{1}{\Lambda_*}\delta\dot{\phi}^2\,{\cal O}\,,
\end{align}
characterized by a new scale $\Lambda_*$, which must be larger than $H$ to ensure that the vertex remains weakly coupled at Hubble crossing.
\begin{figure}
    \centering
\includegraphics[scale=0.8]{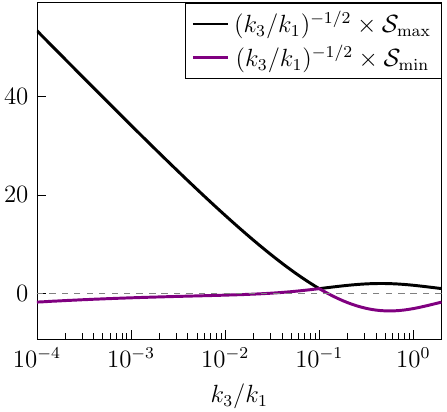}
    \caption{Bounds on the bispectrum shape $\tilde{S}$ (defined by Eq.~\eqref{newshape}) in the Type II setup.}
    \label{fig:SmaxSmin2}
\end{figure}

As we explained in Section \ref{bispecsignsize}, 
the scale $\Lambda_*$ is not tied to the size of the linear mixing $\delta\dot{\phi}\,{\cal O}$, which has to be small compared to the free field Lagrangian, at Hubble crossing. As a result, the single-exchange diagram formed by the above vertex and the original linear mixing can be much larger than the one involving the dS invariant cubic vertex $(\partial_\mu \delta\phi)^2\,{\cal O}$, potentially leading to $f^{{\cal O}^2}_{\text{NL}}\gg 1$. For this reason and for the simplicity of the presentation, in this appendix, we only consider the above cubic vertex \eqref{Lambdastartverrex} and neglect the de Sitter invariant one. We also assume that contributions to the bispectrum from inflaton self-interactions—governed by the EFT operators $\dot{\pi}^3$ and $\dot{\pi}(\partial_i \pi)^2$—are negligible compared to those from the single-exchange process discussed above. This assumption is crucial for establishing any bounds on the bispectrum, as the signs of these EFT operators are unknown.

In the current setup, although the coupling to the inflaton sector strongly breaks de Sitter boosts, we still assume that the hidden sector—particularly its two-point function—remains de Sitter invariant at leading order. Therefore, the general structure of the bispectrum remains dictated by the same spectral decomposition as in Eq.~\eqref{spectralrep}. However, the bispectrum template $B(k_i;\mu)$ entering this expression should be modified---it now corresponds to the exchange of a massive field $f$ that is coupled to $\delta\phi$ with the new cubic vertex $\delta\dot{\phi}^2\,f/\Lambda_*$, while the linear mixing $\delta\dot{\phi}f$ remains unchanged. 

Just like the Type I case, the new templates $B(k_i;\mu)$ are also related to the seed four-point function $\cF_4(u,v)$ albeit via a different weight-shifting operator. Specifically, in this case one finds
\begin{align}
\nn
B(k_1,k_2,k_3;\mu) &=\kappa_*\,\hat{{\cal W}}_*(k_1,k_2,k_3;\partial_u)\,\cF_4(u,v)\Big|_{u=\frac{k_3}{k_1+k_2},v=1}\\
\label{WSostar}
&+(t-u)\,\text{channels},
\end{align}
where 
\begin{align}
        \hat{{\cal W}}_*&=\dfrac{-1}{k_1 k_2 k_3 (k_1+k_2)^3}\left(2\partial_u+u\,\partial^2_u\right)\,,
\end{align}
and 
\begin{align}
\label{kappastar}
\kappa_*=8\pi^3\alpha\dfrac{|\dot{\bar{\phi}}|}{\Lambda \,\Lambda_*}\Delta_\zeta^{3}\,.
\end{align}
Notably, unlike the Type I setup, in Type II we cannot establish any definite sign for the bispectrum based on the positivity of $\rho(\mu)$. The reason is that the overall coupling $\kappa_*$ appearing above is not sign definite. In fact, even with a definite sign for $\kappa_*$, it is not possible to put bounds on the sign of the equilateral bispectrum in Type II, because the template $B(1,1,1;\mu)$ changes sign as $\mu$ is varied, see Fig.~\ref{fig:TypeBtemplates}. This is also a roadblock to putting bounds on the bispectrum shape, conventionally normalized at the equilateral configuration. Such object would be unbounded for the same reason. 

Luckily, in the new setup, there are still configurations for which the bispectra corresponding to the exchange of different masses all have the same sign. As an example, one can numerically verify that 
\begin{align}
    B(1,1,0.1;\mu)<0\,\,\,\forall \,\,\,\mu\geq 0\,.
\end{align}
We use this fiducial configuration to normalize the shape and define 
\begin{align}
\label{newshape}
    \tilde{{\cal S}}(k_i)&=
    \dfrac{(k_1k_2k_3)^2}{0.1^2}\dfrac{B(k_1,k_2,k_3)}{B(1,1,0.1)}\\ \nn
    &=\dfrac{(k_1k_2k_3)^2}{0.1^2}\dfrac{\int \mathrm{d}\mu\,\rho(\mu) B(k_i;\mu)}{\int \mathrm{d}\mu\,\rho(\mu) B(1,1,0.1;\mu)}\,,
\end{align}
which is expected to be bounded from above and below. Akin to the bispectrum shape in Type I, at a fixed momentum configuration, the shape $\tilde{{\cal S}}(k_i)$ is also optimized by a spectral density that sharply picks at a particular mass index $\mu=\mu_{\text{max,min}}$, which we find numerically. In other words, we find the optimal values of the ansatz
\begin{align}
\label{newansatz}
    \tilde{{\cal S}}(k_i;\mu)=\dfrac{(k_1 k_2 k_3)^2}{0.1^2}\,\dfrac{B(k_i;\mu)}{B(1,1,0.1;\mu)}\,,
\end{align}
by varying the mass across the entire principal series, i.e. $\mu\in [0,\infty)$ (see the results in Fig.~\ref{fig:SmaxSmin2}). As shown in Fig.~\ref{fig:Opm2}, unlike the bounds obtained in the Type I setup, the maximum value of $\tilde{{\cal S}}$ is not set by the bispectrum associated with the exchange of the lightest field (i.e. $\mu = 0$). Instead, for configurations with $k_3/k_1 \gtrsim 0.1$, we find that $\mu_{\text{max}} \approx 1.9$ while $\mu_{\text{min}} = 0$. For more squeezed triangles, however, $\mu_{\text{max}}$ decreases to zero, while $\mu_{\text{min}}$ varies from 0 in the squeezed limit up to $\approx 1.6$ as $k_3/k_1$ approaches 0.1.

In addition to the possibilities mentioned in Section \ref{violation}, violation of the Type II bounds might also be attributed to the EFT operators $\dot{\pi}^3$ and $\dot{\pi}(\partial_i \pi)^2$, whose contributions we have neglected. This contrasts with the Type I bounds, which cannot be violated by the leading order inflaton self-interaction induced by the sign-definite operator $(\partial \phi)^4$.

\bibliographystyle{apsrev4-2}
	
\bibliography{refs}

\begin{thebibliography}{103}%
\makeatletter
\providecommand \@ifxundefined [1]{%
 \@ifx{#1\undefined}
}%
\providecommand \@ifnum [1]{%
 \ifnum #1\expandafter \@firstoftwo
 \else \expandafter \@secondoftwo
 \fi
}%
\providecommand \@ifx [1]{%
 \ifx #1\expandafter \@firstoftwo
 \else \expandafter \@secondoftwo
 \fi
}%
\providecommand \natexlab [1]{#1}%
\providecommand \enquote  [1]{``#1''}%
\providecommand \bibnamefont  [1]{#1}%
\providecommand \bibfnamefont [1]{#1}%
\providecommand \citenamefont [1]{#1}%
\providecommand \href@noop [0]{\@secondoftwo}%
\providecommand \href [0]{\begingroup \@sanitize@url \@href}%
\providecommand \@href[1]{\@@startlink{#1}\@@href}%
\providecommand \@@href[1]{\endgroup#1\@@endlink}%
\providecommand \@sanitize@url [0]{\catcode `\\12\catcode `\$12\catcode
  `\&12\catcode `\#12\catcode `\^12\catcode `\_12\catcode `\%12\relax}%
\providecommand \@@startlink[1]{}%
\providecommand \@@endlink[0]{}%
\providecommand \url  [0]{\begingroup\@sanitize@url \@url }%
\providecommand \@url [1]{\endgroup\@href {#1}{\urlprefix }}%
\providecommand \urlprefix  [0]{URL }%
\providecommand \Eprint [0]{\href }%
\providecommand \doibase [0]{https://doi.org/}%
\providecommand \selectlanguage [0]{\@gobble}%
\providecommand \bibinfo  [0]{\@secondoftwo}%
\providecommand \bibfield  [0]{\@secondoftwo}%
\providecommand \translation [1]{[#1]}%
\providecommand \BibitemOpen [0]{}%
\providecommand \bibitemStop [0]{}%
\providecommand \bibitemNoStop [0]{.\EOS\space}%
\providecommand \EOS [0]{\spacefactor3000\relax}%
\providecommand \BibitemShut  [1]{\csname bibitem#1\endcsname}%
\let\auto@bib@innerbib\@empty
\bibitem [{\citenamefont {Akrami}\ \emph {et~al.}(2020)\citenamefont {Akrami}
  \emph {et~al.}}]{Planck:2018jri}%
  \BibitemOpen
  \bibfield  {author} {\bibinfo {author} {\bibfnamefont {Y.}~\bibnamefont
  {Akrami}} \emph {et~al.} (\bibinfo {collaboration} {Planck}),\ }\href
  {https://doi.org/10.1051/0004-6361/201833887} {\bibfield  {journal} {\bibinfo
   {journal} {Astron. Astrophys.}\ }\textbf {\bibinfo {volume} {641}},\
  \bibinfo {pages} {A10} (\bibinfo {year} {2020})},\ \Eprint
  {https://arxiv.org/abs/1807.06211} {arXiv:1807.06211 [astro-ph.CO]}
  \BibitemShut {NoStop}%
\bibitem [{\citenamefont {Maldacena}(2003)}]{Maldacena:2002vr}%
  \BibitemOpen
  \bibfield  {author} {\bibinfo {author} {\bibfnamefont {J.~M.}\ \bibnamefont
  {Maldacena}},\ }\href {https://doi.org/10.1088/1126-6708/2003/05/013}
  {\bibfield  {journal} {\bibinfo  {journal} {JHEP}\ }\textbf {\bibinfo
  {volume} {05}},\ \bibinfo {pages} {013}},\ \Eprint
  {https://arxiv.org/abs/astro-ph/0210603} {arXiv:astro-ph/0210603}
  \BibitemShut {NoStop}%
\bibitem [{\citenamefont {Mu\~noz}\ \emph {et~al.}(2015)\citenamefont
  {Mu\~noz}, \citenamefont {Ali-Ha\"\i{}moud},\ and\ \citenamefont
  {Kamionkowski}}]{Munoz:2015eqa}%
  \BibitemOpen
  \bibfield  {author} {\bibinfo {author} {\bibfnamefont {J.~B.}\ \bibnamefont
  {Mu\~noz}}, \bibinfo {author} {\bibfnamefont {Y.}~\bibnamefont
  {Ali-Ha\"\i{}moud}},\ and\ \bibinfo {author} {\bibfnamefont {M.}~\bibnamefont
  {Kamionkowski}},\ }\href {https://doi.org/10.1103/PhysRevD.92.083508}
  {\bibfield  {journal} {\bibinfo  {journal} {Phys. Rev. D}\ }\textbf {\bibinfo
  {volume} {92}},\ \bibinfo {pages} {083508} (\bibinfo {year} {2015})},\
  \Eprint {https://arxiv.org/abs/1506.04152} {arXiv:1506.04152 [astro-ph.CO]}
  \BibitemShut {NoStop}%
\bibitem [{\citenamefont {Baumann}\ and\ \citenamefont
  {McAllister}(2015)}]{Baumann:2014nda}%
  \BibitemOpen
  \bibfield  {author} {\bibinfo {author} {\bibfnamefont {D.}~\bibnamefont
  {Baumann}}\ and\ \bibinfo {author} {\bibfnamefont {L.}~\bibnamefont
  {McAllister}},\ }\href
  {https://inspirehep.net/record/1289899/files/arXiv:1404.2601.pdf} {\emph
  {\bibinfo {title} {{Inflation and String Theory}}}}\ (\bibinfo  {publisher}
  {Cambridge University Press},\ \bibinfo {year} {2015})\ \Eprint
  {https://arxiv.org/abs/1404.2601} {arXiv:1404.2601 [hep-th]} \BibitemShut
  {NoStop}%
\bibitem [{\citenamefont {Ach\'ucarro}\ \emph {et~al.}(2022)\citenamefont
  {Ach\'ucarro} \emph {et~al.}}]{Achucarro:2022qrl}%
  \BibitemOpen
  \bibfield  {author} {\bibinfo {author} {\bibfnamefont {A.}~\bibnamefont
  {Ach\'ucarro}} \emph {et~al.},\ }\href@noop {} {\  (\bibinfo {year}
  {2022})},\ \Eprint {https://arxiv.org/abs/2203.08128} {arXiv:2203.08128
  [astro-ph.CO]} \BibitemShut {NoStop}%
\bibitem [{\citenamefont {Alishahiha}\ \emph {et~al.}(2004)\citenamefont
  {Alishahiha}, \citenamefont {Silverstein},\ and\ \citenamefont
  {Tong}}]{Alishahiha:2004eh}%
  \BibitemOpen
  \bibfield  {author} {\bibinfo {author} {\bibfnamefont {M.}~\bibnamefont
  {Alishahiha}}, \bibinfo {author} {\bibfnamefont {E.}~\bibnamefont
  {Silverstein}},\ and\ \bibinfo {author} {\bibfnamefont {D.}~\bibnamefont
  {Tong}},\ }\href {https://doi.org/10.1103/PhysRevD.70.123505} {\bibfield
  {journal} {\bibinfo  {journal} {Phys. Rev.}\ }\textbf {\bibinfo {volume}
  {D70}},\ \bibinfo {pages} {123505} (\bibinfo {year} {2004})},\ \Eprint
  {https://arxiv.org/abs/hep-th/0404084} {arXiv:hep-th/0404084} \BibitemShut
  {NoStop}%
\bibitem [{\citenamefont {Tolley}\ and\ \citenamefont
  {Wyman}(2010)}]{Tolley:2009fg}%
  \BibitemOpen
  \bibfield  {author} {\bibinfo {author} {\bibfnamefont {A.~J.}\ \bibnamefont
  {Tolley}}\ and\ \bibinfo {author} {\bibfnamefont {M.}~\bibnamefont {Wyman}},\
  }\href {https://doi.org/10.1103/PhysRevD.81.043502} {\bibfield  {journal}
  {\bibinfo  {journal} {Phys. Rev.}\ }\textbf {\bibinfo {volume} {D81}},\
  \bibinfo {pages} {043502} (\bibinfo {year} {2010})},\ \Eprint
  {https://arxiv.org/abs/0910.1853} {arXiv:0910.1853 [hep-th]} \BibitemShut
  {NoStop}%
\bibitem [{\citenamefont {Baumann}\ and\ \citenamefont
  {Green}(2011)}]{Baumann:2011su}%
  \BibitemOpen
  \bibfield  {author} {\bibinfo {author} {\bibfnamefont {D.}~\bibnamefont
  {Baumann}}\ and\ \bibinfo {author} {\bibfnamefont {D.}~\bibnamefont
  {Green}},\ }\href {https://doi.org/10.1088/1475-7516/2011/09/014} {\bibfield
  {journal} {\bibinfo  {journal} {JCAP}\ }\textbf {\bibinfo {volume} {1109}},\
  \bibinfo {pages} {014}},\ \Eprint {https://arxiv.org/abs/1102.5343}
  {arXiv:1102.5343 [hep-th]} \BibitemShut {NoStop}%
\bibitem [{\citenamefont {Achucarro}\ \emph {et~al.}(2011)\citenamefont
  {Achucarro}, \citenamefont {Gong}, \citenamefont {Hardeman}, \citenamefont
  {Palma},\ and\ \citenamefont {Patil}}]{Achucarro:2010da}%
  \BibitemOpen
  \bibfield  {author} {\bibinfo {author} {\bibfnamefont {A.}~\bibnamefont
  {Achucarro}}, \bibinfo {author} {\bibfnamefont {J.-O.}\ \bibnamefont {Gong}},
  \bibinfo {author} {\bibfnamefont {S.}~\bibnamefont {Hardeman}}, \bibinfo
  {author} {\bibfnamefont {G.~A.}\ \bibnamefont {Palma}},\ and\ \bibinfo
  {author} {\bibfnamefont {S.~P.}\ \bibnamefont {Patil}},\ }\href
  {https://doi.org/10.1088/1475-7516/2011/01/030} {\bibfield  {journal}
  {\bibinfo  {journal} {JCAP}\ }\textbf {\bibinfo {volume} {01}},\ \bibinfo
  {pages} {030}},\ \Eprint {https://arxiv.org/abs/1010.3693} {arXiv:1010.3693
  [hep-ph]} \BibitemShut {NoStop}%
\bibitem [{\citenamefont {Flauger}\ \emph {et~al.}(2017)\citenamefont
  {Flauger}, \citenamefont {Mirbabayi}, \citenamefont {Senatore},\ and\
  \citenamefont {Silverstein}}]{Flauger:2016idt}%
  \BibitemOpen
  \bibfield  {author} {\bibinfo {author} {\bibfnamefont {R.}~\bibnamefont
  {Flauger}}, \bibinfo {author} {\bibfnamefont {M.}~\bibnamefont {Mirbabayi}},
  \bibinfo {author} {\bibfnamefont {L.}~\bibnamefont {Senatore}},\ and\
  \bibinfo {author} {\bibfnamefont {E.}~\bibnamefont {Silverstein}},\ }\href
  {https://doi.org/10.1088/1475-7516/2017/10/058} {\bibfield  {journal}
  {\bibinfo  {journal} {JCAP}\ }\textbf {\bibinfo {volume} {1710}}\bibfield
  {number} {\bibinfo  {number} { (10)},\ \bibinfo {pages} {058}},\ }\Eprint
  {https://arxiv.org/abs/1606.00513} {arXiv:1606.00513 [hep-th]} \BibitemShut
  {NoStop}%
\bibitem [{\citenamefont {Pajer}\ \emph {et~al.}(2024)\citenamefont {Pajer},
  \citenamefont {Wang},\ and\ \citenamefont {Zhang}}]{Pajer:2024ckd}%
  \BibitemOpen
  \bibfield  {author} {\bibinfo {author} {\bibfnamefont {E.}~\bibnamefont
  {Pajer}}, \bibinfo {author} {\bibfnamefont {D.-G.}\ \bibnamefont {Wang}},\
  and\ \bibinfo {author} {\bibfnamefont {B.}~\bibnamefont {Zhang}},\
  }\href@noop {} {\  (\bibinfo {year} {2024})},\ \Eprint
  {https://arxiv.org/abs/2412.05762} {arXiv:2412.05762 [hep-th]} \BibitemShut
  {NoStop}%
\bibitem [{\citenamefont {Cheung}\ \emph {et~al.}(2008)\citenamefont {Cheung},
  \citenamefont {Creminelli}, \citenamefont {Fitzpatrick}, \citenamefont
  {Kaplan},\ and\ \citenamefont {Senatore}}]{Cheung:2007st}%
  \BibitemOpen
  \bibfield  {author} {\bibinfo {author} {\bibfnamefont {C.}~\bibnamefont
  {Cheung}}, \bibinfo {author} {\bibfnamefont {P.}~\bibnamefont {Creminelli}},
  \bibinfo {author} {\bibfnamefont {A.~L.}\ \bibnamefont {Fitzpatrick}},
  \bibinfo {author} {\bibfnamefont {J.}~\bibnamefont {Kaplan}},\ and\ \bibinfo
  {author} {\bibfnamefont {L.}~\bibnamefont {Senatore}},\ }\href
  {https://doi.org/10.1088/1126-6708/2008/03/014} {\bibfield  {journal}
  {\bibinfo  {journal} {JHEP}\ }\textbf {\bibinfo {volume} {03}},\ \bibinfo
  {pages} {014}},\ \Eprint {https://arxiv.org/abs/0709.0293} {arXiv:0709.0293
  [hep-th]} \BibitemShut {NoStop}%
\bibitem [{\citenamefont {Weinberg}(2008)}]{Weinberg:2008hq}%
  \BibitemOpen
  \bibfield  {author} {\bibinfo {author} {\bibfnamefont {S.}~\bibnamefont
  {Weinberg}},\ }\href {https://doi.org/10.1103/PhysRevD.77.123541} {\bibfield
  {journal} {\bibinfo  {journal} {Phys. Rev.}\ }\textbf {\bibinfo {volume}
  {D77}},\ \bibinfo {pages} {123541} (\bibinfo {year} {2008})},\ \Eprint
  {https://arxiv.org/abs/0804.4291} {arXiv:0804.4291 [hep-th]} \BibitemShut
  {NoStop}%
\bibitem [{\citenamefont {Arkani-Hamed}\ and\ \citenamefont
  {Maldacena}(2015)}]{Arkani-Hamed:2015bza}%
  \BibitemOpen
  \bibfield  {author} {\bibinfo {author} {\bibfnamefont {N.}~\bibnamefont
  {Arkani-Hamed}}\ and\ \bibinfo {author} {\bibfnamefont {J.}~\bibnamefont
  {Maldacena}},\ }\href@noop {} {\  (\bibinfo {year} {2015})},\ \Eprint
  {https://arxiv.org/abs/1503.08043} {arXiv:1503.08043 [hep-th]} \BibitemShut
  {NoStop}%
\bibitem [{\citenamefont {Lee}\ \emph {et~al.}(2016)\citenamefont {Lee},
  \citenamefont {Baumann},\ and\ \citenamefont {Pimentel}}]{Lee:2016vti}%
  \BibitemOpen
  \bibfield  {author} {\bibinfo {author} {\bibfnamefont {H.}~\bibnamefont
  {Lee}}, \bibinfo {author} {\bibfnamefont {D.}~\bibnamefont {Baumann}},\ and\
  \bibinfo {author} {\bibfnamefont {G.~L.}\ \bibnamefont {Pimentel}},\ }\href
  {https://doi.org/10.1007/JHEP12(2016)040} {\bibfield  {journal} {\bibinfo
  {journal} {JHEP}\ }\textbf {\bibinfo {volume} {12}},\ \bibinfo {pages}
  {040}},\ \Eprint {https://arxiv.org/abs/1607.03735} {arXiv:1607.03735
  [hep-th]} \BibitemShut {NoStop}%
\bibitem [{\citenamefont {Chen}\ \emph
  {et~al.}(2017{\natexlab{a}})\citenamefont {Chen}, \citenamefont {Wang},\ and\
  \citenamefont {Xianyu}}]{Chen:2016uwp}%
  \BibitemOpen
  \bibfield  {author} {\bibinfo {author} {\bibfnamefont {X.}~\bibnamefont
  {Chen}}, \bibinfo {author} {\bibfnamefont {Y.}~\bibnamefont {Wang}},\ and\
  \bibinfo {author} {\bibfnamefont {Z.-Z.}\ \bibnamefont {Xianyu}},\ }\href
  {https://doi.org/10.1103/PhysRevLett.118.261302} {\bibfield  {journal}
  {\bibinfo  {journal} {Phys. Rev. Lett.}\ }\textbf {\bibinfo {volume} {118}},\
  \bibinfo {pages} {261302} (\bibinfo {year} {2017}{\natexlab{a}})},\ \Eprint
  {https://arxiv.org/abs/1610.06597} {arXiv:1610.06597 [hep-th]} \BibitemShut
  {NoStop}%
\bibitem [{\citenamefont {Chen}\ \emph
  {et~al.}(2017{\natexlab{b}})\citenamefont {Chen}, \citenamefont {Wang},\ and\
  \citenamefont {Xianyu}}]{Chen:2016hrz}%
  \BibitemOpen
  \bibfield  {author} {\bibinfo {author} {\bibfnamefont {X.}~\bibnamefont
  {Chen}}, \bibinfo {author} {\bibfnamefont {Y.}~\bibnamefont {Wang}},\ and\
  \bibinfo {author} {\bibfnamefont {Z.-Z.}\ \bibnamefont {Xianyu}},\ }\href
  {https://doi.org/10.1007/JHEP04(2017)058} {\bibfield  {journal} {\bibinfo
  {journal} {JHEP}\ }\textbf {\bibinfo {volume} {04}},\ \bibinfo {pages}
  {058}},\ \Eprint {https://arxiv.org/abs/1612.08122} {arXiv:1612.08122
  [hep-th]} \BibitemShut {NoStop}%
\bibitem [{\citenamefont {Wang}\ and\ \citenamefont
  {Xianyu}(2020)}]{Wang:2019gbi}%
  \BibitemOpen
  \bibfield  {author} {\bibinfo {author} {\bibfnamefont {L.-T.}\ \bibnamefont
  {Wang}}\ and\ \bibinfo {author} {\bibfnamefont {Z.-Z.}\ \bibnamefont
  {Xianyu}},\ }\href {https://doi.org/10.1007/JHEP02(2020)044} {\bibfield
  {journal} {\bibinfo  {journal} {JHEP}\ }\textbf {\bibinfo {volume} {02}},\
  \bibinfo {pages} {044}},\ \Eprint {https://arxiv.org/abs/1910.12876}
  {arXiv:1910.12876 [hep-ph]} \BibitemShut {NoStop}%
\bibitem [{\citenamefont {Kumar}\ and\ \citenamefont
  {Sundrum}(2020)}]{Kumar:2019ebj}%
  \BibitemOpen
  \bibfield  {author} {\bibinfo {author} {\bibfnamefont {S.}~\bibnamefont
  {Kumar}}\ and\ \bibinfo {author} {\bibfnamefont {R.}~\bibnamefont
  {Sundrum}},\ }\href {https://doi.org/10.1007/JHEP04(2020)077} {\bibfield
  {journal} {\bibinfo  {journal} {JHEP}\ }\textbf {\bibinfo {volume} {04}},\
  \bibinfo {pages} {077}},\ \Eprint {https://arxiv.org/abs/1908.11378}
  {arXiv:1908.11378 [hep-ph]} \BibitemShut {NoStop}%
\bibitem [{\citenamefont {Bodas}\ \emph {et~al.}(2021)\citenamefont {Bodas},
  \citenamefont {Kumar},\ and\ \citenamefont {Sundrum}}]{Bodas:2020yho}%
  \BibitemOpen
  \bibfield  {author} {\bibinfo {author} {\bibfnamefont {A.}~\bibnamefont
  {Bodas}}, \bibinfo {author} {\bibfnamefont {S.}~\bibnamefont {Kumar}},\ and\
  \bibinfo {author} {\bibfnamefont {R.}~\bibnamefont {Sundrum}},\ }\href
  {https://doi.org/10.1007/JHEP02(2021)079} {\bibfield  {journal} {\bibinfo
  {journal} {JHEP}\ }\textbf {\bibinfo {volume} {02}},\ \bibinfo {pages}
  {079}},\ \Eprint {https://arxiv.org/abs/2010.04727} {arXiv:2010.04727
  [hep-ph]} \BibitemShut {NoStop}%
\bibitem [{\citenamefont {Tong}\ \emph {et~al.}(2022)\citenamefont {Tong},
  \citenamefont {Wang},\ and\ \citenamefont {Zhu}}]{Tong:2021wai}%
  \BibitemOpen
  \bibfield  {author} {\bibinfo {author} {\bibfnamefont {X.}~\bibnamefont
  {Tong}}, \bibinfo {author} {\bibfnamefont {Y.}~\bibnamefont {Wang}},\ and\
  \bibinfo {author} {\bibfnamefont {Y.}~\bibnamefont {Zhu}},\ }\href
  {https://doi.org/10.1007/JHEP03(2022)181} {\bibfield  {journal} {\bibinfo
  {journal} {JHEP}\ }\textbf {\bibinfo {volume} {03}},\ \bibinfo {pages}
  {181}},\ \Eprint {https://arxiv.org/abs/2112.03448} {arXiv:2112.03448
  [hep-th]} \BibitemShut {NoStop}%
\bibitem [{\citenamefont {Pinol}\ \emph {et~al.}(2023)\citenamefont {Pinol},
  \citenamefont {Aoki}, \citenamefont {Renaux-Petel},\ and\ \citenamefont
  {Yamaguchi}}]{Pinol:2021aun}%
  \BibitemOpen
  \bibfield  {author} {\bibinfo {author} {\bibfnamefont {L.}~\bibnamefont
  {Pinol}}, \bibinfo {author} {\bibfnamefont {S.}~\bibnamefont {Aoki}},
  \bibinfo {author} {\bibfnamefont {S.}~\bibnamefont {Renaux-Petel}},\ and\
  \bibinfo {author} {\bibfnamefont {M.}~\bibnamefont {Yamaguchi}},\ }\href
  {https://doi.org/10.1103/PhysRevD.107.L021301} {\bibfield  {journal}
  {\bibinfo  {journal} {Phys. Rev. D}\ }\textbf {\bibinfo {volume} {107}},\
  \bibinfo {pages} {L021301} (\bibinfo {year} {2023})},\ \Eprint
  {https://arxiv.org/abs/2112.05710} {arXiv:2112.05710 [hep-th]} \BibitemShut
  {NoStop}%
\bibitem [{\citenamefont {Tong}\ and\ \citenamefont
  {Xianyu}(2022)}]{Tong:2022cdz}%
  \BibitemOpen
  \bibfield  {author} {\bibinfo {author} {\bibfnamefont {X.}~\bibnamefont
  {Tong}}\ and\ \bibinfo {author} {\bibfnamefont {Z.-Z.}\ \bibnamefont
  {Xianyu}},\ }\href {https://doi.org/10.1007/JHEP10(2022)194} {\bibfield
  {journal} {\bibinfo  {journal} {JHEP}\ }\textbf {\bibinfo {volume} {10}},\
  \bibinfo {pages} {194}},\ \Eprint {https://arxiv.org/abs/2203.06349}
  {arXiv:2203.06349 [hep-ph]} \BibitemShut {NoStop}%
\bibitem [{\citenamefont {Hubisz}\ \emph {et~al.}(2025)\citenamefont {Hubisz},
  \citenamefont {Lee}, \citenamefont {Li},\ and\ \citenamefont
  {Sambasivam}}]{Hubisz:2024xnj}%
  \BibitemOpen
  \bibfield  {author} {\bibinfo {author} {\bibfnamefont {J.}~\bibnamefont
  {Hubisz}}, \bibinfo {author} {\bibfnamefont {S.~J.}\ \bibnamefont {Lee}},
  \bibinfo {author} {\bibfnamefont {H.}~\bibnamefont {Li}},\ and\ \bibinfo
  {author} {\bibfnamefont {B.}~\bibnamefont {Sambasivam}},\ }\href
  {https://doi.org/10.1103/PhysRevD.111.023543} {\bibfield  {journal} {\bibinfo
   {journal} {Phys. Rev. D}\ }\textbf {\bibinfo {volume} {111}},\ \bibinfo
  {pages} {023543} (\bibinfo {year} {2025})},\ \Eprint
  {https://arxiv.org/abs/2408.08951} {arXiv:2408.08951 [astro-ph.CO]}
  \BibitemShut {NoStop}%
\bibitem [{\citenamefont {Aoki}(2023)}]{Aoki:2023tjm}%
  \BibitemOpen
  \bibfield  {author} {\bibinfo {author} {\bibfnamefont {S.}~\bibnamefont
  {Aoki}},\ }\href {https://doi.org/10.1088/1475-7516/2023/04/002} {\bibfield
  {journal} {\bibinfo  {journal} {JCAP}\ }\textbf {\bibinfo {volume} {04}},\
  \bibinfo {pages} {002}},\ \Eprint {https://arxiv.org/abs/2301.07920}
  {arXiv:2301.07920 [hep-th]} \BibitemShut {NoStop}%
\bibitem [{\citenamefont {de~Rham}\ \emph {et~al.}(2017)\citenamefont
  {de~Rham}, \citenamefont {Melville}, \citenamefont {Tolley},\ and\
  \citenamefont {Zhou}}]{deRham:2017avq}%
  \BibitemOpen
  \bibfield  {author} {\bibinfo {author} {\bibfnamefont {C.}~\bibnamefont
  {de~Rham}}, \bibinfo {author} {\bibfnamefont {S.}~\bibnamefont {Melville}},
  \bibinfo {author} {\bibfnamefont {A.~J.}\ \bibnamefont {Tolley}},\ and\
  \bibinfo {author} {\bibfnamefont {S.-Y.}\ \bibnamefont {Zhou}},\ }\href
  {https://doi.org/10.1103/PhysRevD.96.081702} {\bibfield  {journal} {\bibinfo
  {journal} {Phys. Rev. D}\ }\textbf {\bibinfo {volume} {96}},\ \bibinfo
  {pages} {081702} (\bibinfo {year} {2017})},\ \Eprint
  {https://arxiv.org/abs/1702.06134} {arXiv:1702.06134 [hep-th]} \BibitemShut
  {NoStop}%
\bibitem [{\citenamefont {de~Rham}\ \emph {et~al.}(2018)\citenamefont
  {de~Rham}, \citenamefont {Melville}, \citenamefont {Tolley},\ and\
  \citenamefont {Zhou}}]{deRham:2017zjm}%
  \BibitemOpen
  \bibfield  {author} {\bibinfo {author} {\bibfnamefont {C.}~\bibnamefont
  {de~Rham}}, \bibinfo {author} {\bibfnamefont {S.}~\bibnamefont {Melville}},
  \bibinfo {author} {\bibfnamefont {A.~J.}\ \bibnamefont {Tolley}},\ and\
  \bibinfo {author} {\bibfnamefont {S.-Y.}\ \bibnamefont {Zhou}},\ }\href
  {https://doi.org/10.1007/JHEP03(2018)011} {\bibfield  {journal} {\bibinfo
  {journal} {JHEP}\ }\textbf {\bibinfo {volume} {03}},\ \bibinfo {pages}
  {011}},\ \Eprint {https://arxiv.org/abs/1706.02712} {arXiv:1706.02712
  [hep-th]} \BibitemShut {NoStop}%
\bibitem [{\citenamefont {de~Rham}\ \emph {et~al.}(2019)\citenamefont
  {de~Rham}, \citenamefont {Melville}, \citenamefont {Tolley},\ and\
  \citenamefont {Zhou}}]{deRham:2018qqo}%
  \BibitemOpen
  \bibfield  {author} {\bibinfo {author} {\bibfnamefont {C.}~\bibnamefont
  {de~Rham}}, \bibinfo {author} {\bibfnamefont {S.}~\bibnamefont {Melville}},
  \bibinfo {author} {\bibfnamefont {A.~J.}\ \bibnamefont {Tolley}},\ and\
  \bibinfo {author} {\bibfnamefont {S.-Y.}\ \bibnamefont {Zhou}},\ }\href
  {https://doi.org/10.1007/JHEP03(2019)182} {\bibfield  {journal} {\bibinfo
  {journal} {JHEP}\ }\textbf {\bibinfo {volume} {03}},\ \bibinfo {pages}
  {182}},\ \Eprint {https://arxiv.org/abs/1804.10624} {arXiv:1804.10624
  [hep-th]} \BibitemShut {NoStop}%
\bibitem [{\citenamefont {Paulos}\ \emph {et~al.}(2019)\citenamefont {Paulos},
  \citenamefont {Penedones}, \citenamefont {Toledo}, \citenamefont {van Rees},\
  and\ \citenamefont {Vieira}}]{Paulos:2017fhb}%
  \BibitemOpen
  \bibfield  {author} {\bibinfo {author} {\bibfnamefont {M.~F.}\ \bibnamefont
  {Paulos}}, \bibinfo {author} {\bibfnamefont {J.}~\bibnamefont {Penedones}},
  \bibinfo {author} {\bibfnamefont {J.}~\bibnamefont {Toledo}}, \bibinfo
  {author} {\bibfnamefont {B.~C.}\ \bibnamefont {van Rees}},\ and\ \bibinfo
  {author} {\bibfnamefont {P.}~\bibnamefont {Vieira}},\ }\href
  {https://doi.org/10.1007/JHEP12(2019)040} {\bibfield  {journal} {\bibinfo
  {journal} {JHEP}\ }\textbf {\bibinfo {volume} {12}},\ \bibinfo {pages}
  {040}},\ \Eprint {https://arxiv.org/abs/1708.06765} {arXiv:1708.06765
  [hep-th]} \BibitemShut {NoStop}%
\bibitem [{\citenamefont {Arkani-Hamed}\ \emph {et~al.}(2021)\citenamefont
  {Arkani-Hamed}, \citenamefont {Huang},\ and\ \citenamefont
  {Huang}}]{Arkani-Hamed:2020blm}%
  \BibitemOpen
  \bibfield  {author} {\bibinfo {author} {\bibfnamefont {N.}~\bibnamefont
  {Arkani-Hamed}}, \bibinfo {author} {\bibfnamefont {T.-C.}\ \bibnamefont
  {Huang}},\ and\ \bibinfo {author} {\bibfnamefont {Y.-t.}\ \bibnamefont
  {Huang}},\ }\href {https://doi.org/10.1007/JHEP05(2021)259} {\bibfield
  {journal} {\bibinfo  {journal} {JHEP}\ }\textbf {\bibinfo {volume} {05}},\
  \bibinfo {pages} {259}},\ \Eprint {https://arxiv.org/abs/2012.15849}
  {arXiv:2012.15849 [hep-th]} \BibitemShut {NoStop}%
\bibitem [{\citenamefont {Tolley}\ \emph {et~al.}(2021)\citenamefont {Tolley},
  \citenamefont {Wang},\ and\ \citenamefont {Zhou}}]{Tolley:2020gtv}%
  \BibitemOpen
  \bibfield  {author} {\bibinfo {author} {\bibfnamefont {A.~J.}\ \bibnamefont
  {Tolley}}, \bibinfo {author} {\bibfnamefont {Z.-Y.}\ \bibnamefont {Wang}},\
  and\ \bibinfo {author} {\bibfnamefont {S.-Y.}\ \bibnamefont {Zhou}},\ }\href
  {https://doi.org/10.1007/JHEP05(2021)255} {\bibfield  {journal} {\bibinfo
  {journal} {JHEP}\ }\textbf {\bibinfo {volume} {05}},\ \bibinfo {pages}
  {255}},\ \Eprint {https://arxiv.org/abs/2011.02400} {arXiv:2011.02400
  [hep-th]} \BibitemShut {NoStop}%
\bibitem [{\citenamefont {Caron-Huot}\ and\ \citenamefont
  {Van~Duong}(2021)}]{Caron-Huot:2020cmc}%
  \BibitemOpen
  \bibfield  {author} {\bibinfo {author} {\bibfnamefont {S.}~\bibnamefont
  {Caron-Huot}}\ and\ \bibinfo {author} {\bibfnamefont {V.}~\bibnamefont
  {Van~Duong}},\ }\href {https://doi.org/10.1007/JHEP05(2021)280} {\bibfield
  {journal} {\bibinfo  {journal} {JHEP}\ }\textbf {\bibinfo {volume} {05}},\
  \bibinfo {pages} {280}},\ \Eprint {https://arxiv.org/abs/2011.02957}
  {arXiv:2011.02957 [hep-th]} \BibitemShut {NoStop}%
\bibitem [{\citenamefont {Bellazzini}\ \emph {et~al.}(2021)\citenamefont
  {Bellazzini}, \citenamefont {Elias~Mir\'o}, \citenamefont {Rattazzi},
  \citenamefont {Riembau},\ and\ \citenamefont {Riva}}]{Bellazzini:2020cot}%
  \BibitemOpen
  \bibfield  {author} {\bibinfo {author} {\bibfnamefont {B.}~\bibnamefont
  {Bellazzini}}, \bibinfo {author} {\bibfnamefont {J.}~\bibnamefont
  {Elias~Mir\'o}}, \bibinfo {author} {\bibfnamefont {R.}~\bibnamefont
  {Rattazzi}}, \bibinfo {author} {\bibfnamefont {M.}~\bibnamefont {Riembau}},\
  and\ \bibinfo {author} {\bibfnamefont {F.}~\bibnamefont {Riva}},\ }\href
  {https://doi.org/10.1103/PhysRevD.104.036006} {\bibfield  {journal} {\bibinfo
   {journal} {Phys. Rev. D}\ }\textbf {\bibinfo {volume} {104}},\ \bibinfo
  {pages} {036006} (\bibinfo {year} {2021})},\ \Eprint
  {https://arxiv.org/abs/2011.00037} {arXiv:2011.00037 [hep-th]} \BibitemShut
  {NoStop}%
\bibitem [{\citenamefont {Caron-Huot}\ \emph {et~al.}(2021)\citenamefont
  {Caron-Huot}, \citenamefont {Mazac}, \citenamefont {Rastelli},\ and\
  \citenamefont {Simmons-Duffin}}]{Caron-Huot:2021rmr}%
  \BibitemOpen
  \bibfield  {author} {\bibinfo {author} {\bibfnamefont {S.}~\bibnamefont
  {Caron-Huot}}, \bibinfo {author} {\bibfnamefont {D.}~\bibnamefont {Mazac}},
  \bibinfo {author} {\bibfnamefont {L.}~\bibnamefont {Rastelli}},\ and\
  \bibinfo {author} {\bibfnamefont {D.}~\bibnamefont {Simmons-Duffin}},\ }\href
  {https://doi.org/10.1007/JHEP07(2021)110} {\bibfield  {journal} {\bibinfo
  {journal} {JHEP}\ }\textbf {\bibinfo {volume} {07}},\ \bibinfo {pages}
  {110}},\ \Eprint {https://arxiv.org/abs/2102.08951} {arXiv:2102.08951
  [hep-th]} \BibitemShut {NoStop}%
\bibitem [{\citenamefont {Guerrieri}\ and\ \citenamefont
  {Sever}(2021)}]{Guerrieri:2021tak}%
  \BibitemOpen
  \bibfield  {author} {\bibinfo {author} {\bibfnamefont {A.}~\bibnamefont
  {Guerrieri}}\ and\ \bibinfo {author} {\bibfnamefont {A.}~\bibnamefont
  {Sever}},\ }\href {https://doi.org/10.1103/PhysRevLett.127.251601} {\bibfield
   {journal} {\bibinfo  {journal} {Phys. Rev. Lett.}\ }\textbf {\bibinfo
  {volume} {127}},\ \bibinfo {pages} {251601} (\bibinfo {year} {2021})},\
  \Eprint {https://arxiv.org/abs/2106.10257} {arXiv:2106.10257 [hep-th]}
  \BibitemShut {NoStop}%
\bibitem [{\citenamefont {de~Rham}\ \emph {et~al.}(2022)\citenamefont
  {de~Rham}, \citenamefont {Kundu}, \citenamefont {Reece}, \citenamefont
  {Tolley},\ and\ \citenamefont {Zhou}}]{deRham:2022hpx}%
  \BibitemOpen
  \bibfield  {author} {\bibinfo {author} {\bibfnamefont {C.}~\bibnamefont
  {de~Rham}}, \bibinfo {author} {\bibfnamefont {S.}~\bibnamefont {Kundu}},
  \bibinfo {author} {\bibfnamefont {M.}~\bibnamefont {Reece}}, \bibinfo
  {author} {\bibfnamefont {A.~J.}\ \bibnamefont {Tolley}},\ and\ \bibinfo
  {author} {\bibfnamefont {S.-Y.}\ \bibnamefont {Zhou}},\ }in\ \href@noop {}
  {\emph {\bibinfo {booktitle} {{Snowmass 2021}}}}\ (\bibinfo {year} {2022})\
  \Eprint {https://arxiv.org/abs/2203.06805} {arXiv:2203.06805 [hep-th]}
  \BibitemShut {NoStop}%
\bibitem [{\citenamefont {Kruczenski}\ \emph {et~al.}(2022)\citenamefont
  {Kruczenski}, \citenamefont {Penedones},\ and\ \citenamefont {van
  Rees}}]{Kruczenski:2022lot}%
  \BibitemOpen
  \bibfield  {author} {\bibinfo {author} {\bibfnamefont {M.}~\bibnamefont
  {Kruczenski}}, \bibinfo {author} {\bibfnamefont {J.}~\bibnamefont
  {Penedones}},\ and\ \bibinfo {author} {\bibfnamefont {B.~C.}\ \bibnamefont
  {van Rees}},\ }\href@noop {} {\  (\bibinfo {year} {2022})},\ \Eprint
  {https://arxiv.org/abs/2203.02421} {arXiv:2203.02421 [hep-th]} \BibitemShut
  {NoStop}%
\bibitem [{\citenamefont {de~Rham}\ and\ \citenamefont
  {Melville}(2017)}]{deRham:2017aoj}%
  \BibitemOpen
  \bibfield  {author} {\bibinfo {author} {\bibfnamefont {C.}~\bibnamefont
  {de~Rham}}\ and\ \bibinfo {author} {\bibfnamefont {S.}~\bibnamefont
  {Melville}},\ }\href {https://doi.org/10.1103/PhysRevD.95.123523} {\bibfield
  {journal} {\bibinfo  {journal} {Phys. Rev. D}\ }\textbf {\bibinfo {volume}
  {95}},\ \bibinfo {pages} {123523} (\bibinfo {year} {2017})},\ \Eprint
  {https://arxiv.org/abs/1703.00025} {arXiv:1703.00025 [hep-th]} \BibitemShut
  {NoStop}%
\bibitem [{\citenamefont {Grall}\ and\ \citenamefont
  {Melville}(2020)}]{Grall:2020tqc}%
  \BibitemOpen
  \bibfield  {author} {\bibinfo {author} {\bibfnamefont {T.}~\bibnamefont
  {Grall}}\ and\ \bibinfo {author} {\bibfnamefont {S.}~\bibnamefont
  {Melville}},\ }\href {https://doi.org/10.1088/1475-7516/2020/09/017}
  {\bibfield  {journal} {\bibinfo  {journal} {JCAP}\ }\textbf {\bibinfo
  {volume} {09}},\ \bibinfo {pages} {017}},\ \Eprint
  {https://arxiv.org/abs/2005.02366} {arXiv:2005.02366 [gr-qc]} \BibitemShut
  {NoStop}%
\bibitem [{\citenamefont {Grall}\ and\ \citenamefont
  {Melville}(2022)}]{Grall:2021xxm}%
  \BibitemOpen
  \bibfield  {author} {\bibinfo {author} {\bibfnamefont {T.}~\bibnamefont
  {Grall}}\ and\ \bibinfo {author} {\bibfnamefont {S.}~\bibnamefont
  {Melville}},\ }\href {https://doi.org/10.1103/PhysRevD.105.L121301}
  {\bibfield  {journal} {\bibinfo  {journal} {Phys. Rev. D}\ }\textbf {\bibinfo
  {volume} {105}},\ \bibinfo {pages} {L121301} (\bibinfo {year} {2022})},\
  \Eprint {https://arxiv.org/abs/2102.05683} {arXiv:2102.05683 [hep-th]}
  \BibitemShut {NoStop}%
\bibitem [{\citenamefont {Grall}(2023)}]{Grall:2022pad}%
  \BibitemOpen
  \bibfield  {author} {\bibinfo {author} {\bibfnamefont {T.}~\bibnamefont
  {Grall}},\ }\emph {\bibinfo {title} {{Symmetries, unitarity and positivity of
  cosmological effective field theories}}},\ \href
  {https://doi.org/10.17863/CAM.95639} {Ph.D. thesis},\ \bibinfo  {school}
  {Department of Applied Mathematics And Theoretical Physics, Cambridge U.,
  University of Cambridge} (\bibinfo {year} {2023})\BibitemShut {NoStop}%
\bibitem [{\citenamefont {Stefanyszyn}\ \emph {et~al.}(2024)\citenamefont
  {Stefanyszyn}, \citenamefont {Tong},\ and\ \citenamefont
  {Zhu}}]{Stefanyszyn:2023qov}%
  \BibitemOpen
  \bibfield  {author} {\bibinfo {author} {\bibfnamefont {D.}~\bibnamefont
  {Stefanyszyn}}, \bibinfo {author} {\bibfnamefont {X.}~\bibnamefont {Tong}},\
  and\ \bibinfo {author} {\bibfnamefont {Y.}~\bibnamefont {Zhu}},\ }\href
  {https://doi.org/10.1007/JHEP05(2024)196} {\bibfield  {journal} {\bibinfo
  {journal} {JHEP}\ }\textbf {\bibinfo {volume} {05}},\ \bibinfo {pages}
  {196}},\ \Eprint {https://arxiv.org/abs/2309.07769} {arXiv:2309.07769
  [hep-th]} \BibitemShut {NoStop}%
\bibitem [{\citenamefont {Grall}\ \emph
  {et~al.}(2020{\natexlab{a}})\citenamefont {Grall}, \citenamefont {Jazayeri},\
  and\ \citenamefont {Pajer}}]{Grall:2019qof}%
  \BibitemOpen
  \bibfield  {author} {\bibinfo {author} {\bibfnamefont {T.}~\bibnamefont
  {Grall}}, \bibinfo {author} {\bibfnamefont {S.}~\bibnamefont {Jazayeri}},\
  and\ \bibinfo {author} {\bibfnamefont {E.}~\bibnamefont {Pajer}},\ }\href
  {https://doi.org/10.1088/1475-7516/2020/05/031} {\bibfield  {journal}
  {\bibinfo  {journal} {JCAP}\ }\textbf {\bibinfo {volume} {05}},\ \bibinfo
  {pages} {031}},\ \Eprint {https://arxiv.org/abs/1909.04622} {arXiv:1909.04622
  [hep-th]} \BibitemShut {NoStop}%
\bibitem [{\citenamefont {Grall}\ \emph
  {et~al.}(2020{\natexlab{b}})\citenamefont {Grall}, \citenamefont {Jazayeri},\
  and\ \citenamefont {Stefanyszyn}}]{Grall:2020ibl}%
  \BibitemOpen
  \bibfield  {author} {\bibinfo {author} {\bibfnamefont {T.}~\bibnamefont
  {Grall}}, \bibinfo {author} {\bibfnamefont {S.}~\bibnamefont {Jazayeri}},\
  and\ \bibinfo {author} {\bibfnamefont {D.}~\bibnamefont {Stefanyszyn}},\
  }\href {https://doi.org/10.1007/JHEP11(2020)097} {\bibfield  {journal}
  {\bibinfo  {journal} {JHEP}\ }\textbf {\bibinfo {volume} {11}},\ \bibinfo
  {pages} {097}},\ \Eprint {https://arxiv.org/abs/2005.12937} {arXiv:2005.12937
  [hep-th]} \BibitemShut {NoStop}%
\bibitem [{\citenamefont {Agui~Salcedo}\ and\ \citenamefont
  {Melville}(2023)}]{AguiSalcedo:2023nds}%
  \BibitemOpen
  \bibfield  {author} {\bibinfo {author} {\bibfnamefont {S.}~\bibnamefont
  {Agui~Salcedo}}\ and\ \bibinfo {author} {\bibfnamefont {S.}~\bibnamefont
  {Melville}},\ }\href {https://doi.org/10.1007/JHEP12(2023)076} {\bibfield
  {journal} {\bibinfo  {journal} {JHEP}\ }\textbf {\bibinfo {volume} {12}},\
  \bibinfo {pages} {076}},\ \Eprint {https://arxiv.org/abs/2308.00680}
  {arXiv:2308.00680 [hep-th]} \BibitemShut {NoStop}%
\bibitem [{\citenamefont {Melville}\ and\ \citenamefont
  {Pimentel}(2024{\natexlab{a}})}]{Melville:2023kgd}%
  \BibitemOpen
  \bibfield  {author} {\bibinfo {author} {\bibfnamefont {S.}~\bibnamefont
  {Melville}}\ and\ \bibinfo {author} {\bibfnamefont {G.~L.}\ \bibnamefont
  {Pimentel}},\ }\href {https://doi.org/10.1103/PhysRevD.110.103530} {\bibfield
   {journal} {\bibinfo  {journal} {Phys. Rev. D}\ }\textbf {\bibinfo {volume}
  {110}},\ \bibinfo {pages} {103530} (\bibinfo {year} {2024}{\natexlab{a}})},\
  \Eprint {https://arxiv.org/abs/2309.07092} {arXiv:2309.07092 [hep-th]}
  \BibitemShut {NoStop}%
\bibitem [{\citenamefont {Melville}\ and\ \citenamefont
  {Pimentel}(2024{\natexlab{b}})}]{Melville:2024ove}%
  \BibitemOpen
  \bibfield  {author} {\bibinfo {author} {\bibfnamefont {S.}~\bibnamefont
  {Melville}}\ and\ \bibinfo {author} {\bibfnamefont {G.~L.}\ \bibnamefont
  {Pimentel}},\ }\href {https://doi.org/10.1007/JHEP08(2024)211} {\bibfield
  {journal} {\bibinfo  {journal} {JHEP}\ }\textbf {\bibinfo {volume} {08}},\
  \bibinfo {pages} {211}},\ \Eprint {https://arxiv.org/abs/2404.05712}
  {arXiv:2404.05712 [hep-th]} \BibitemShut {NoStop}%
\bibitem [{\citenamefont {Cespedes}\ and\ \citenamefont
  {Jazayeri}(2025)}]{Cespedes:2025dnq}%
  \BibitemOpen
  \bibfield  {author} {\bibinfo {author} {\bibfnamefont {S.}~\bibnamefont
  {Cespedes}}\ and\ \bibinfo {author} {\bibfnamefont {S.}~\bibnamefont
  {Jazayeri}},\ }\href@noop {} {\  (\bibinfo {year} {2025})},\ \Eprint
  {https://arxiv.org/abs/2501.02119} {arXiv:2501.02119 [hep-th]} \BibitemShut
  {NoStop}%
\bibitem [{\citenamefont {Green}\ \emph {et~al.}(2013)\citenamefont {Green},
  \citenamefont {Lewandowski}, \citenamefont {Senatore}, \citenamefont
  {Silverstein},\ and\ \citenamefont {Zaldarriaga}}]{Green:2013rd}%
  \BibitemOpen
  \bibfield  {author} {\bibinfo {author} {\bibfnamefont {D.}~\bibnamefont
  {Green}}, \bibinfo {author} {\bibfnamefont {M.}~\bibnamefont {Lewandowski}},
  \bibinfo {author} {\bibfnamefont {L.}~\bibnamefont {Senatore}}, \bibinfo
  {author} {\bibfnamefont {E.}~\bibnamefont {Silverstein}},\ and\ \bibinfo
  {author} {\bibfnamefont {M.}~\bibnamefont {Zaldarriaga}},\ }\href
  {https://doi.org/10.1007/JHEP10(2013)171} {\bibfield  {journal} {\bibinfo
  {journal} {JHEP}\ }\textbf {\bibinfo {volume} {10}},\ \bibinfo {pages}
  {171}},\ \Eprint {https://arxiv.org/abs/1301.2630} {arXiv:1301.2630 [hep-th]}
  \BibitemShut {NoStop}%
\bibitem [{\citenamefont {Pimentel}\ and\ \citenamefont
  {Yang}(2025)}]{Pimentel:2025rds}%
  \BibitemOpen
  \bibfield  {author} {\bibinfo {author} {\bibfnamefont {G.~L.}\ \bibnamefont
  {Pimentel}}\ and\ \bibinfo {author} {\bibfnamefont {C.}~\bibnamefont
  {Yang}},\ }\href@noop {} {\  (\bibinfo {year} {2025})},\ \Eprint
  {https://arxiv.org/abs/2503.17840} {arXiv:2503.17840 [hep-th]} \BibitemShut
  {NoStop}%
\bibitem [{\citenamefont {Hogervorst}\ \emph {et~al.}(2023)\citenamefont
  {Hogervorst}, \citenamefont {Penedones},\ and\ \citenamefont
  {Vaziri}}]{Hogervorst:2021uvp}%
  \BibitemOpen
  \bibfield  {author} {\bibinfo {author} {\bibfnamefont {M.}~\bibnamefont
  {Hogervorst}}, \bibinfo {author} {\bibfnamefont {J.~a.}\ \bibnamefont
  {Penedones}},\ and\ \bibinfo {author} {\bibfnamefont {K.~S.}\ \bibnamefont
  {Vaziri}},\ }\href {https://doi.org/10.1007/JHEP02(2023)162} {\bibfield
  {journal} {\bibinfo  {journal} {JHEP}\ }\textbf {\bibinfo {volume} {02}},\
  \bibinfo {pages} {162}},\ \Eprint {https://arxiv.org/abs/2107.13871}
  {arXiv:2107.13871 [hep-th]} \BibitemShut {NoStop}%
\bibitem [{\citenamefont {Di~Pietro}\ \emph {et~al.}(2021)\citenamefont
  {Di~Pietro}, \citenamefont {Gorbenko},\ and\ \citenamefont
  {Komatsu}}]{DiPietro:2021sjt}%
  \BibitemOpen
  \bibfield  {author} {\bibinfo {author} {\bibfnamefont {L.}~\bibnamefont
  {Di~Pietro}}, \bibinfo {author} {\bibfnamefont {V.}~\bibnamefont
  {Gorbenko}},\ and\ \bibinfo {author} {\bibfnamefont {S.}~\bibnamefont
  {Komatsu}},\ }\href@noop {} {\  (\bibinfo {year} {2021})},\ \Eprint
  {https://arxiv.org/abs/2108.01695} {arXiv:2108.01695 [hep-th]} \BibitemShut
  {NoStop}%
\bibitem [{\citenamefont {Di~Pietro}\ \emph {et~al.}(2023)\citenamefont
  {Di~Pietro}, \citenamefont {Gorbenko},\ and\ \citenamefont
  {Komatsu}}]{DiPietro:2023inn}%
  \BibitemOpen
  \bibfield  {author} {\bibinfo {author} {\bibfnamefont {L.}~\bibnamefont
  {Di~Pietro}}, \bibinfo {author} {\bibfnamefont {V.}~\bibnamefont
  {Gorbenko}},\ and\ \bibinfo {author} {\bibfnamefont {S.}~\bibnamefont
  {Komatsu}},\ }\href@noop {} {\  (\bibinfo {year} {2023})},\ \Eprint
  {https://arxiv.org/abs/2312.17195} {arXiv:2312.17195 [hep-th]} \BibitemShut
  {NoStop}%
\bibitem [{\citenamefont {Penedones}\ \emph {et~al.}(2023)\citenamefont
  {Penedones}, \citenamefont {Salehi~Vaziri},\ and\ \citenamefont
  {Sun}}]{Penedones:2023uqc}%
  \BibitemOpen
  \bibfield  {author} {\bibinfo {author} {\bibfnamefont {J.}~\bibnamefont
  {Penedones}}, \bibinfo {author} {\bibfnamefont {K.}~\bibnamefont
  {Salehi~Vaziri}},\ and\ \bibinfo {author} {\bibfnamefont {Z.}~\bibnamefont
  {Sun}},\ }\href@noop {} {\  (\bibinfo {year} {2023})},\ \Eprint
  {https://arxiv.org/abs/2301.04146} {arXiv:2301.04146 [hep-th]} \BibitemShut
  {NoStop}%
\bibitem [{\citenamefont {Loparco}\ \emph {et~al.}(2023)\citenamefont
  {Loparco}, \citenamefont {Penedones}, \citenamefont {Salehi~Vaziri},\ and\
  \citenamefont {Sun}}]{Loparco:2023rug}%
  \BibitemOpen
  \bibfield  {author} {\bibinfo {author} {\bibfnamefont {M.}~\bibnamefont
  {Loparco}}, \bibinfo {author} {\bibfnamefont {J.}~\bibnamefont {Penedones}},
  \bibinfo {author} {\bibfnamefont {K.}~\bibnamefont {Salehi~Vaziri}},\ and\
  \bibinfo {author} {\bibfnamefont {Z.}~\bibnamefont {Sun}},\ }\href
  {https://doi.org/10.1007/JHEP12(2023)159} {\bibfield  {journal} {\bibinfo
  {journal} {JHEP}\ }\textbf {\bibinfo {volume} {12}},\ \bibinfo {pages}
  {159}},\ \Eprint {https://arxiv.org/abs/2306.00090} {arXiv:2306.00090
  [hep-th]} \BibitemShut {NoStop}%
\bibitem [{\citenamefont {Loparco}\ \emph {et~al.}(2025)\citenamefont
  {Loparco}, \citenamefont {Penedones},\ and\ \citenamefont
  {Ulrich}}]{Loparco:2025azm}%
  \BibitemOpen
  \bibfield  {author} {\bibinfo {author} {\bibfnamefont {M.}~\bibnamefont
  {Loparco}}, \bibinfo {author} {\bibfnamefont {J.}~\bibnamefont {Penedones}},\
  and\ \bibinfo {author} {\bibfnamefont {Y.}~\bibnamefont {Ulrich}},\
  }\href@noop {} {\  (\bibinfo {year} {2025})},\ \Eprint
  {https://arxiv.org/abs/2505.00761} {arXiv:2505.00761 [hep-th]} \BibitemShut
  {NoStop}%
\bibitem [{\citenamefont {Creminelli}(2003)}]{Creminelli:2003iq}%
  \BibitemOpen
  \bibfield  {author} {\bibinfo {author} {\bibfnamefont {P.}~\bibnamefont
  {Creminelli}},\ }\href {https://doi.org/10.1088/1475-7516/2003/10/003}
  {\bibfield  {journal} {\bibinfo  {journal} {JCAP}\ }\textbf {\bibinfo
  {volume} {10}},\ \bibinfo {pages} {003}},\ \Eprint
  {https://arxiv.org/abs/astro-ph/0306122} {arXiv:astro-ph/0306122}
  \BibitemShut {NoStop}%
\bibitem [{\citenamefont {Pham}\ and\ \citenamefont
  {Truong}(1985)}]{Pham:1985cr}%
  \BibitemOpen
  \bibfield  {author} {\bibinfo {author} {\bibfnamefont {T.~N.}\ \bibnamefont
  {Pham}}\ and\ \bibinfo {author} {\bibfnamefont {T.~N.}\ \bibnamefont
  {Truong}},\ }\href {https://doi.org/10.1103/PhysRevD.31.3027} {\bibfield
  {journal} {\bibinfo  {journal} {Phys. Rev. D}\ }\textbf {\bibinfo {volume}
  {31}},\ \bibinfo {pages} {3027} (\bibinfo {year} {1985})}\BibitemShut
  {NoStop}%
\bibitem [{\citenamefont {Adams}\ \emph {et~al.}(2006)\citenamefont {Adams},
  \citenamefont {Arkani-Hamed}, \citenamefont {Dubovsky}, \citenamefont
  {Nicolis},\ and\ \citenamefont {Rattazzi}}]{Adams:2006sv}%
  \BibitemOpen
  \bibfield  {author} {\bibinfo {author} {\bibfnamefont {A.}~\bibnamefont
  {Adams}}, \bibinfo {author} {\bibfnamefont {N.}~\bibnamefont {Arkani-Hamed}},
  \bibinfo {author} {\bibfnamefont {S.}~\bibnamefont {Dubovsky}}, \bibinfo
  {author} {\bibfnamefont {A.}~\bibnamefont {Nicolis}},\ and\ \bibinfo {author}
  {\bibfnamefont {R.}~\bibnamefont {Rattazzi}},\ }\href
  {https://doi.org/10.1088/1126-6708/2006/10/014} {\bibfield  {journal}
  {\bibinfo  {journal} {JHEP}\ }\textbf {\bibinfo {volume} {10}},\ \bibinfo
  {pages} {014}},\ \Eprint {https://arxiv.org/abs/hep-th/0602178}
  {arXiv:hep-th/0602178} \BibitemShut {NoStop}%
\bibitem [{\citenamefont {Marolf}\ and\ \citenamefont
  {Morrison}(2010)}]{Marolf:2010zp}%
  \BibitemOpen
  \bibfield  {author} {\bibinfo {author} {\bibfnamefont {D.}~\bibnamefont
  {Marolf}}\ and\ \bibinfo {author} {\bibfnamefont {I.~A.}\ \bibnamefont
  {Morrison}},\ }\href {https://doi.org/10.1103/PhysRevD.82.105032} {\bibfield
  {journal} {\bibinfo  {journal} {Phys. Rev. D}\ }\textbf {\bibinfo {volume}
  {82}},\ \bibinfo {pages} {105032} (\bibinfo {year} {2010})},\ \Eprint
  {https://arxiv.org/abs/1006.0035} {arXiv:1006.0035 [gr-qc]} \BibitemShut
  {NoStop}%
\bibitem [{\citenamefont {Xianyu}\ and\ \citenamefont
  {Zhang}(2023)}]{Xianyu:2022jwk}%
  \BibitemOpen
  \bibfield  {author} {\bibinfo {author} {\bibfnamefont {Z.-Z.}\ \bibnamefont
  {Xianyu}}\ and\ \bibinfo {author} {\bibfnamefont {H.}~\bibnamefont {Zhang}},\
  }\href {https://doi.org/10.1007/JHEP04(2023)103} {\bibfield  {journal}
  {\bibinfo  {journal} {JHEP}\ }\textbf {\bibinfo {volume} {04}},\ \bibinfo
  {pages} {103}},\ \Eprint {https://arxiv.org/abs/2211.03810} {arXiv:2211.03810
  [hep-th]} \BibitemShut {NoStop}%
\bibitem [{\citenamefont {Baumann}\ \emph {et~al.}(2024)\citenamefont
  {Baumann}, \citenamefont {Green}, \citenamefont {Joyce}, \citenamefont
  {Pajer}, \citenamefont {Pimentel}, \citenamefont {Sleight},\ and\
  \citenamefont {Taronna}}]{Baumann:2022jpr}%
  \BibitemOpen
  \bibfield  {author} {\bibinfo {author} {\bibfnamefont {D.}~\bibnamefont
  {Baumann}}, \bibinfo {author} {\bibfnamefont {D.}~\bibnamefont {Green}},
  \bibinfo {author} {\bibfnamefont {A.}~\bibnamefont {Joyce}}, \bibinfo
  {author} {\bibfnamefont {E.}~\bibnamefont {Pajer}}, \bibinfo {author}
  {\bibfnamefont {G.~L.}\ \bibnamefont {Pimentel}}, \bibinfo {author}
  {\bibfnamefont {C.}~\bibnamefont {Sleight}},\ and\ \bibinfo {author}
  {\bibfnamefont {M.}~\bibnamefont {Taronna}},\ }\href
  {https://doi.org/10.21468/SciPostPhysCommRep.1} {\bibfield  {journal}
  {\bibinfo  {journal} {SciPost Phys. Comm. Rep.}\ }\textbf {\bibinfo {volume}
  {2024}},\ \bibinfo {pages} {1} (\bibinfo {year} {2024})},\ \Eprint
  {https://arxiv.org/abs/2203.08121} {arXiv:2203.08121 [hep-th]} \BibitemShut
  {NoStop}%
\bibitem [{\citenamefont {Arkani-Hamed}\ \emph {et~al.}(2020)\citenamefont
  {Arkani-Hamed}, \citenamefont {Baumann}, \citenamefont {Lee},\ and\
  \citenamefont {Pimentel}}]{Arkani-Hamed:2018kmz}%
  \BibitemOpen
  \bibfield  {author} {\bibinfo {author} {\bibfnamefont {N.}~\bibnamefont
  {Arkani-Hamed}}, \bibinfo {author} {\bibfnamefont {D.}~\bibnamefont
  {Baumann}}, \bibinfo {author} {\bibfnamefont {H.}~\bibnamefont {Lee}},\ and\
  \bibinfo {author} {\bibfnamefont {G.~L.}\ \bibnamefont {Pimentel}},\ }\href
  {https://doi.org/10.1007/JHEP04(2020)105} {\bibfield  {journal} {\bibinfo
  {journal} {JHEP}\ }\textbf {\bibinfo {volume} {04}},\ \bibinfo {pages}
  {105}},\ \Eprint {https://arxiv.org/abs/1811.00024} {arXiv:1811.00024
  [hep-th]} \BibitemShut {NoStop}%
\bibitem [{\citenamefont {Baumann}\ \emph {et~al.}(2020)\citenamefont
  {Baumann}, \citenamefont {Duaso~Pueyo}, \citenamefont {Joyce}, \citenamefont
  {Lee},\ and\ \citenamefont {Pimentel}}]{Baumann:2019oyu}%
  \BibitemOpen
  \bibfield  {author} {\bibinfo {author} {\bibfnamefont {D.}~\bibnamefont
  {Baumann}}, \bibinfo {author} {\bibfnamefont {C.}~\bibnamefont
  {Duaso~Pueyo}}, \bibinfo {author} {\bibfnamefont {A.}~\bibnamefont {Joyce}},
  \bibinfo {author} {\bibfnamefont {H.}~\bibnamefont {Lee}},\ and\ \bibinfo
  {author} {\bibfnamefont {G.~L.}\ \bibnamefont {Pimentel}},\ }\href
  {https://doi.org/10.1007/JHEP12(2020)204} {\bibfield  {journal} {\bibinfo
  {journal} {JHEP}\ }\textbf {\bibinfo {volume} {12}},\ \bibinfo {pages}
  {204}},\ \Eprint {https://arxiv.org/abs/1910.14051} {arXiv:1910.14051
  [hep-th]} \BibitemShut {NoStop}%
\bibitem [{\citenamefont {Jazayeri}\ and\ \citenamefont
  {Renaux-Petel}(2022)}]{Jazayeri:2022kjy}%
  \BibitemOpen
  \bibfield  {author} {\bibinfo {author} {\bibfnamefont {S.}~\bibnamefont
  {Jazayeri}}\ and\ \bibinfo {author} {\bibfnamefont {S.}~\bibnamefont
  {Renaux-Petel}},\ }\href {https://doi.org/10.1007/JHEP12(2022)137} {\bibfield
   {journal} {\bibinfo  {journal} {JHEP}\ }\textbf {\bibinfo {volume} {12}},\
  \bibinfo {pages} {137}},\ \Eprint {https://arxiv.org/abs/2205.10340}
  {arXiv:2205.10340 [hep-th]} \BibitemShut {NoStop}%
\bibitem [{\citenamefont {Qin}\ and\ \citenamefont
  {Xianyu}(2023)}]{Qin:2023ejc}%
  \BibitemOpen
  \bibfield  {author} {\bibinfo {author} {\bibfnamefont {Z.}~\bibnamefont
  {Qin}}\ and\ \bibinfo {author} {\bibfnamefont {Z.-Z.}\ \bibnamefont
  {Xianyu}},\ }\href {https://doi.org/10.1007/JHEP07(2023)001} {\bibfield
  {journal} {\bibinfo  {journal} {JHEP}\ }\textbf {\bibinfo {volume} {07}},\
  \bibinfo {pages} {001}},\ \Eprint {https://arxiv.org/abs/2301.07047}
  {arXiv:2301.07047 [hep-th]} \BibitemShut {NoStop}%
\bibitem [{\citenamefont {Senatore}\ and\ \citenamefont
  {Zaldarriaga}(2012)}]{Senatore:2010wk}%
  \BibitemOpen
  \bibfield  {author} {\bibinfo {author} {\bibfnamefont {L.}~\bibnamefont
  {Senatore}}\ and\ \bibinfo {author} {\bibfnamefont {M.}~\bibnamefont
  {Zaldarriaga}},\ }\href {https://doi.org/10.1007/JHEP04(2012)024} {\bibfield
  {journal} {\bibinfo  {journal} {JHEP}\ }\textbf {\bibinfo {volume} {04}},\
  \bibinfo {pages} {024}},\ \Eprint {https://arxiv.org/abs/1009.2093}
  {arXiv:1009.2093 [hep-th]} \BibitemShut {NoStop}%
\bibitem [{\citenamefont {Noumi}\ \emph {et~al.}(2013)\citenamefont {Noumi},
  \citenamefont {Yamaguchi},\ and\ \citenamefont {Yokoyama}}]{Noumi:2012vr}%
  \BibitemOpen
  \bibfield  {author} {\bibinfo {author} {\bibfnamefont {T.}~\bibnamefont
  {Noumi}}, \bibinfo {author} {\bibfnamefont {M.}~\bibnamefont {Yamaguchi}},\
  and\ \bibinfo {author} {\bibfnamefont {D.}~\bibnamefont {Yokoyama}},\ }\href
  {https://doi.org/10.1007/JHEP06(2013)051} {\bibfield  {journal} {\bibinfo
  {journal} {JHEP}\ }\textbf {\bibinfo {volume} {06}},\ \bibinfo {pages}
  {051}},\ \Eprint {https://arxiv.org/abs/1211.1624} {arXiv:1211.1624 [hep-th]}
  \BibitemShut {NoStop}%
\bibitem [{\citenamefont {Garcia-Saenz}\ \emph {et~al.}(2020)\citenamefont
  {Garcia-Saenz}, \citenamefont {Pinol},\ and\ \citenamefont
  {Renaux-Petel}}]{Garcia-Saenz:2019njm}%
  \BibitemOpen
  \bibfield  {author} {\bibinfo {author} {\bibfnamefont {S.}~\bibnamefont
  {Garcia-Saenz}}, \bibinfo {author} {\bibfnamefont {L.}~\bibnamefont
  {Pinol}},\ and\ \bibinfo {author} {\bibfnamefont {S.}~\bibnamefont
  {Renaux-Petel}},\ }\href {https://doi.org/10.1007/JHEP01(2020)073} {\bibfield
   {journal} {\bibinfo  {journal} {JHEP}\ }\textbf {\bibinfo {volume} {01}},\
  \bibinfo {pages} {073}},\ \Eprint {https://arxiv.org/abs/1907.10403}
  {arXiv:1907.10403 [hep-th]} \BibitemShut {NoStop}%
\bibitem [{\citenamefont {Carrillo~Gonz\'alez}\ and\ \citenamefont
  {C\'espedes}(2025)}]{CarrilloGonzalez:2025fqq}%
  \BibitemOpen
  \bibfield  {author} {\bibinfo {author} {\bibfnamefont {M.}~\bibnamefont
  {Carrillo~Gonz\'alez}}\ and\ \bibinfo {author} {\bibfnamefont
  {S.}~\bibnamefont {C\'espedes}},\ }\href@noop {} {\  (\bibinfo {year}
  {2025})},\ \Eprint {https://arxiv.org/abs/2502.19477} {arXiv:2502.19477
  [hep-th]} \BibitemShut {NoStop}%
\bibitem [{\citenamefont {Hui}\ \emph {et~al.}(2025)\citenamefont {Hui},
  \citenamefont {Nicolis}, \citenamefont {Podo},\ and\ \citenamefont
  {Zhou}}]{Hui:2025aja}%
  \BibitemOpen
  \bibfield  {author} {\bibinfo {author} {\bibfnamefont {L.}~\bibnamefont
  {Hui}}, \bibinfo {author} {\bibfnamefont {A.}~\bibnamefont {Nicolis}},
  \bibinfo {author} {\bibfnamefont {A.}~\bibnamefont {Podo}},\ and\ \bibinfo
  {author} {\bibfnamefont {S.}~\bibnamefont {Zhou}},\ }\href@noop {} {\
  (\bibinfo {year} {2025})},\ \Eprint {https://arxiv.org/abs/2502.04215}
  {arXiv:2502.04215 [hep-th]} \BibitemShut {NoStop}%
\bibitem [{\citenamefont {Creminelli}\ \emph {et~al.}(2024)\citenamefont
  {Creminelli}, \citenamefont {Janssen}, \citenamefont {Salehian},\ and\
  \citenamefont {Senatore}}]{Creminelli:2024lhd}%
  \BibitemOpen
  \bibfield  {author} {\bibinfo {author} {\bibfnamefont {P.}~\bibnamefont
  {Creminelli}}, \bibinfo {author} {\bibfnamefont {O.}~\bibnamefont {Janssen}},
  \bibinfo {author} {\bibfnamefont {B.}~\bibnamefont {Salehian}},\ and\
  \bibinfo {author} {\bibfnamefont {L.}~\bibnamefont {Senatore}},\ }\href
  {https://doi.org/10.1007/JHEP08(2024)066} {\bibfield  {journal} {\bibinfo
  {journal} {JHEP}\ }\textbf {\bibinfo {volume} {08}},\ \bibinfo {pages}
  {066}},\ \Eprint {https://arxiv.org/abs/2405.09614} {arXiv:2405.09614
  [hep-th]} \BibitemShut {NoStop}%
\bibitem [{\citenamefont
  {Carrillo~Gonz\'alez}(2024)}]{CarrilloGonzalez:2023emp}%
  \BibitemOpen
  \bibfield  {author} {\bibinfo {author} {\bibfnamefont {M.}~\bibnamefont
  {Carrillo~Gonz\'alez}},\ }\href {https://doi.org/10.1103/PhysRevD.109.085008}
  {\bibfield  {journal} {\bibinfo  {journal} {Phys. Rev. D}\ }\textbf {\bibinfo
  {volume} {109}},\ \bibinfo {pages} {085008} (\bibinfo {year} {2024})},\
  \Eprint {https://arxiv.org/abs/2312.07651} {arXiv:2312.07651 [hep-th]}
  \BibitemShut {NoStop}%
\bibitem [{\citenamefont {Creminelli}\ \emph {et~al.}(2022)\citenamefont
  {Creminelli}, \citenamefont {Janssen},\ and\ \citenamefont
  {Senatore}}]{Creminelli:2022onn}%
  \BibitemOpen
  \bibfield  {author} {\bibinfo {author} {\bibfnamefont {P.}~\bibnamefont
  {Creminelli}}, \bibinfo {author} {\bibfnamefont {O.}~\bibnamefont
  {Janssen}},\ and\ \bibinfo {author} {\bibfnamefont {L.}~\bibnamefont
  {Senatore}},\ }\href {https://doi.org/10.1007/JHEP09(2022)201} {\bibfield
  {journal} {\bibinfo  {journal} {JHEP}\ }\textbf {\bibinfo {volume} {09}},\
  \bibinfo {pages} {201}},\ \Eprint {https://arxiv.org/abs/2207.14224}
  {arXiv:2207.14224 [hep-th]} \BibitemShut {NoStop}%
\bibitem [{\citenamefont {de~Rham}\ \emph {et~al.}(2021)\citenamefont
  {de~Rham}, \citenamefont {Melville},\ and\ \citenamefont
  {Noller}}]{deRham:2021fpu}%
  \BibitemOpen
  \bibfield  {author} {\bibinfo {author} {\bibfnamefont {C.}~\bibnamefont
  {de~Rham}}, \bibinfo {author} {\bibfnamefont {S.}~\bibnamefont {Melville}},\
  and\ \bibinfo {author} {\bibfnamefont {J.}~\bibnamefont {Noller}},\ }\href
  {https://doi.org/10.1088/1475-7516/2021/08/018} {\bibfield  {journal}
  {\bibinfo  {journal} {JCAP}\ }\textbf {\bibinfo {volume} {08}},\ \bibinfo
  {pages} {018}},\ \Eprint {https://arxiv.org/abs/2103.06855} {arXiv:2103.06855
  [astro-ph.CO]} \BibitemShut {NoStop}%
\bibitem [{\citenamefont {de~Rham}\ and\ \citenamefont
  {Tolley}(2020)}]{deRham:2020zyh}%
  \BibitemOpen
  \bibfield  {author} {\bibinfo {author} {\bibfnamefont {C.}~\bibnamefont
  {de~Rham}}\ and\ \bibinfo {author} {\bibfnamefont {A.~J.}\ \bibnamefont
  {Tolley}},\ }\href {https://doi.org/10.1103/PhysRevD.102.084048} {\bibfield
  {journal} {\bibinfo  {journal} {Phys. Rev. D}\ }\textbf {\bibinfo {volume}
  {102}},\ \bibinfo {pages} {084048} (\bibinfo {year} {2020})},\ \Eprint
  {https://arxiv.org/abs/2007.01847} {arXiv:2007.01847 [hep-th]} \BibitemShut
  {NoStop}%
\bibitem [{\citenamefont {Strominger}(2001)}]{Strominger:2001pn}%
  \BibitemOpen
  \bibfield  {author} {\bibinfo {author} {\bibfnamefont {A.}~\bibnamefont
  {Strominger}},\ }\href {https://doi.org/10.1088/1126-6708/2001/10/034}
  {\bibfield  {journal} {\bibinfo  {journal} {JHEP}\ }\textbf {\bibinfo
  {volume} {10}},\ \bibinfo {pages} {034}},\ \Eprint
  {https://arxiv.org/abs/hep-th/0106113} {arXiv:hep-th/0106113} \BibitemShut
  {NoStop}%
\bibitem [{\citenamefont {Khoury}\ \emph {et~al.}(2002)\citenamefont {Khoury},
  \citenamefont {Ovrut}, \citenamefont {Seiberg}, \citenamefont {Steinhardt},\
  and\ \citenamefont {Turok}}]{Khoury:2001bz}%
  \BibitemOpen
  \bibfield  {author} {\bibinfo {author} {\bibfnamefont {J.}~\bibnamefont
  {Khoury}}, \bibinfo {author} {\bibfnamefont {B.~A.}\ \bibnamefont {Ovrut}},
  \bibinfo {author} {\bibfnamefont {N.}~\bibnamefont {Seiberg}}, \bibinfo
  {author} {\bibfnamefont {P.~J.}\ \bibnamefont {Steinhardt}},\ and\ \bibinfo
  {author} {\bibfnamefont {N.}~\bibnamefont {Turok}},\ }\href
  {https://doi.org/10.1103/PhysRevD.65.086007} {\bibfield  {journal} {\bibinfo
  {journal} {Phys. Rev. D}\ }\textbf {\bibinfo {volume} {65}},\ \bibinfo
  {pages} {086007} (\bibinfo {year} {2002})},\ \Eprint
  {https://arxiv.org/abs/hep-th/0108187} {arXiv:hep-th/0108187} \BibitemShut
  {NoStop}%
\bibitem [{\citenamefont {Khoury}\ \emph {et~al.}(2001)\citenamefont {Khoury},
  \citenamefont {Ovrut}, \citenamefont {Steinhardt},\ and\ \citenamefont
  {Turok}}]{Khoury:2001wf}%
  \BibitemOpen
  \bibfield  {author} {\bibinfo {author} {\bibfnamefont {J.}~\bibnamefont
  {Khoury}}, \bibinfo {author} {\bibfnamefont {B.~A.}\ \bibnamefont {Ovrut}},
  \bibinfo {author} {\bibfnamefont {P.~J.}\ \bibnamefont {Steinhardt}},\ and\
  \bibinfo {author} {\bibfnamefont {N.}~\bibnamefont {Turok}},\ }\href
  {https://doi.org/10.1103/PhysRevD.64.123522} {\bibfield  {journal} {\bibinfo
  {journal} {Phys. Rev. D}\ }\textbf {\bibinfo {volume} {64}},\ \bibinfo
  {pages} {123522} (\bibinfo {year} {2001})},\ \Eprint
  {https://arxiv.org/abs/hep-th/0103239} {arXiv:hep-th/0103239} \BibitemShut
  {NoStop}%
\bibitem [{\citenamefont {Gasperini}\ and\ \citenamefont
  {Veneziano}(2003)}]{Gasperini:2002bn}%
  \BibitemOpen
  \bibfield  {author} {\bibinfo {author} {\bibfnamefont {M.}~\bibnamefont
  {Gasperini}}\ and\ \bibinfo {author} {\bibfnamefont {G.}~\bibnamefont
  {Veneziano}},\ }\href {https://doi.org/10.1016/S0370-1573(02)00389-7}
  {\bibfield  {journal} {\bibinfo  {journal} {Phys. Rept.}\ }\textbf {\bibinfo
  {volume} {373}},\ \bibinfo {pages} {1} (\bibinfo {year} {2003})},\ \Eprint
  {https://arxiv.org/abs/hep-th/0207130} {arXiv:hep-th/0207130} \BibitemShut
  {NoStop}%
\bibitem [{\citenamefont {Steinhardt}\ and\ \citenamefont
  {Turok}(2002)}]{Steinhardt:2001st}%
  \BibitemOpen
  \bibfield  {author} {\bibinfo {author} {\bibfnamefont {P.~J.}\ \bibnamefont
  {Steinhardt}}\ and\ \bibinfo {author} {\bibfnamefont {N.}~\bibnamefont
  {Turok}},\ }\href {https://doi.org/10.1103/PhysRevD.65.126003} {\bibfield
  {journal} {\bibinfo  {journal} {Phys. Rev. D}\ }\textbf {\bibinfo {volume}
  {65}},\ \bibinfo {pages} {126003} (\bibinfo {year} {2002})},\ \Eprint
  {https://arxiv.org/abs/hep-th/0111098} {arXiv:hep-th/0111098} \BibitemShut
  {NoStop}%
\bibitem [{\citenamefont {Steinhardt}\ \emph {et~al.}(2002)\citenamefont
  {Steinhardt}, \citenamefont {Turok},\ and\ \citenamefont
  {Turok}}]{Steinhardt:2002ih}%
  \BibitemOpen
  \bibfield  {author} {\bibinfo {author} {\bibfnamefont {P.~J.}\ \bibnamefont
  {Steinhardt}}, \bibinfo {author} {\bibfnamefont {N.}~\bibnamefont {Turok}},\
  and\ \bibinfo {author} {\bibfnamefont {N.}~\bibnamefont {Turok}},\ }\href
  {https://doi.org/10.1126/science.1070462} {\bibfield  {journal} {\bibinfo
  {journal} {Science}\ }\textbf {\bibinfo {volume} {296}},\ \bibinfo {pages}
  {1436} (\bibinfo {year} {2002})},\ \Eprint
  {https://arxiv.org/abs/hep-th/0111030} {arXiv:hep-th/0111030} \BibitemShut
  {NoStop}%
\bibitem [{\citenamefont {Khoury}\ \emph {et~al.}(2004)\citenamefont {Khoury},
  \citenamefont {Steinhardt},\ and\ \citenamefont {Turok}}]{Khoury:2003rt}%
  \BibitemOpen
  \bibfield  {author} {\bibinfo {author} {\bibfnamefont {J.}~\bibnamefont
  {Khoury}}, \bibinfo {author} {\bibfnamefont {P.~J.}\ \bibnamefont
  {Steinhardt}},\ and\ \bibinfo {author} {\bibfnamefont {N.}~\bibnamefont
  {Turok}},\ }\href {https://doi.org/10.1103/PhysRevLett.92.031302} {\bibfield
  {journal} {\bibinfo  {journal} {Phys. Rev. Lett.}\ }\textbf {\bibinfo
  {volume} {92}},\ \bibinfo {pages} {031302} (\bibinfo {year} {2004})},\
  \Eprint {https://arxiv.org/abs/hep-th/0307132} {arXiv:hep-th/0307132}
  \BibitemShut {NoStop}%
\bibitem [{\citenamefont {Tolley}\ and\ \citenamefont
  {Wesley}(2007)}]{Tolley:2007nq}%
  \BibitemOpen
  \bibfield  {author} {\bibinfo {author} {\bibfnamefont {A.~J.}\ \bibnamefont
  {Tolley}}\ and\ \bibinfo {author} {\bibfnamefont {D.~H.}\ \bibnamefont
  {Wesley}},\ }\href {https://doi.org/10.1088/1475-7516/2007/05/006} {\bibfield
   {journal} {\bibinfo  {journal} {JCAP}\ }\textbf {\bibinfo {volume} {05}},\
  \bibinfo {pages} {006}},\ \Eprint {https://arxiv.org/abs/hep-th/0703101}
  {arXiv:hep-th/0703101} \BibitemShut {NoStop}%
\bibitem [{\citenamefont {Creminelli}\ and\ \citenamefont
  {Senatore}(2007)}]{Creminelli:2007aq}%
  \BibitemOpen
  \bibfield  {author} {\bibinfo {author} {\bibfnamefont {P.}~\bibnamefont
  {Creminelli}}\ and\ \bibinfo {author} {\bibfnamefont {L.}~\bibnamefont
  {Senatore}},\ }\href {https://doi.org/10.1088/1475-7516/2007/11/010}
  {\bibfield  {journal} {\bibinfo  {journal} {JCAP}\ }\textbf {\bibinfo
  {volume} {11}},\ \bibinfo {pages} {010}},\ \Eprint
  {https://arxiv.org/abs/hep-th/0702165} {arXiv:hep-th/0702165} \BibitemShut
  {NoStop}%
\bibitem [{\citenamefont {Cai}\ \emph {et~al.}(2012)\citenamefont {Cai},
  \citenamefont {Easson},\ and\ \citenamefont {Brandenberger}}]{Cai:2012va}%
  \BibitemOpen
  \bibfield  {author} {\bibinfo {author} {\bibfnamefont {Y.-F.}\ \bibnamefont
  {Cai}}, \bibinfo {author} {\bibfnamefont {D.~A.}\ \bibnamefont {Easson}},\
  and\ \bibinfo {author} {\bibfnamefont {R.}~\bibnamefont {Brandenberger}},\
  }\href {https://doi.org/10.1088/1475-7516/2012/08/020} {\bibfield  {journal}
  {\bibinfo  {journal} {JCAP}\ }\textbf {\bibinfo {volume} {08}},\ \bibinfo
  {pages} {020}},\ \Eprint {https://arxiv.org/abs/1206.2382} {arXiv:1206.2382
  [hep-th]} \BibitemShut {NoStop}%
\bibitem [{\citenamefont {Brandenberger}(2012)}]{Brandenberger:2012zb}%
  \BibitemOpen
  \bibfield  {author} {\bibinfo {author} {\bibfnamefont {R.~H.}\ \bibnamefont
  {Brandenberger}},\ }\href@noop {} {\  (\bibinfo {year} {2012})},\ \Eprint
  {https://arxiv.org/abs/1206.4196} {arXiv:1206.4196 [astro-ph.CO]}
  \BibitemShut {NoStop}%
\bibitem [{\citenamefont {Battefeld}\ and\ \citenamefont
  {Peter}(2015)}]{Battefeld:2014uga}%
  \BibitemOpen
  \bibfield  {author} {\bibinfo {author} {\bibfnamefont {D.}~\bibnamefont
  {Battefeld}}\ and\ \bibinfo {author} {\bibfnamefont {P.}~\bibnamefont
  {Peter}},\ }\href {https://doi.org/10.1016/j.physrep.2014.12.004} {\bibfield
  {journal} {\bibinfo  {journal} {Phys. Rept.}\ }\textbf {\bibinfo {volume}
  {571}},\ \bibinfo {pages} {1} (\bibinfo {year} {2015})},\ \Eprint
  {https://arxiv.org/abs/1406.2790} {arXiv:1406.2790 [astro-ph.CO]}
  \BibitemShut {NoStop}%
\bibitem [{\citenamefont {Xue}\ \emph {et~al.}(2013)\citenamefont {Xue},
  \citenamefont {Garfinkle}, \citenamefont {Pretorius},\ and\ \citenamefont
  {Steinhardt}}]{Xue:2013bva}%
  \BibitemOpen
  \bibfield  {author} {\bibinfo {author} {\bibfnamefont {B.}~\bibnamefont
  {Xue}}, \bibinfo {author} {\bibfnamefont {D.}~\bibnamefont {Garfinkle}},
  \bibinfo {author} {\bibfnamefont {F.}~\bibnamefont {Pretorius}},\ and\
  \bibinfo {author} {\bibfnamefont {P.~J.}\ \bibnamefont {Steinhardt}},\ }\href
  {https://doi.org/10.1103/PhysRevD.88.083509} {\bibfield  {journal} {\bibinfo
  {journal} {Phys. Rev. D}\ }\textbf {\bibinfo {volume} {88}},\ \bibinfo
  {pages} {083509} (\bibinfo {year} {2013})},\ \Eprint
  {https://arxiv.org/abs/1308.3044} {arXiv:1308.3044 [gr-qc]} \BibitemShut
  {NoStop}%
\bibitem [{\citenamefont {Brandenberger}\ and\ \citenamefont
  {Peter}(2017)}]{Brandenberger:2016vhg}%
  \BibitemOpen
  \bibfield  {author} {\bibinfo {author} {\bibfnamefont {R.}~\bibnamefont
  {Brandenberger}}\ and\ \bibinfo {author} {\bibfnamefont {P.}~\bibnamefont
  {Peter}},\ }\href {https://doi.org/10.1007/s10701-016-0057-0} {\bibfield
  {journal} {\bibinfo  {journal} {Found. Phys.}\ }\textbf {\bibinfo {volume}
  {47}},\ \bibinfo {pages} {797} (\bibinfo {year} {2017})},\ \Eprint
  {https://arxiv.org/abs/1603.05834} {arXiv:1603.05834 [hep-th]} \BibitemShut
  {NoStop}%
\bibitem [{\citenamefont {Ijjas}\ and\ \citenamefont
  {Steinhardt}(2017)}]{Ijjas:2016vtq}%
  \BibitemOpen
  \bibfield  {author} {\bibinfo {author} {\bibfnamefont {A.}~\bibnamefont
  {Ijjas}}\ and\ \bibinfo {author} {\bibfnamefont {P.~J.}\ \bibnamefont
  {Steinhardt}},\ }\href {https://doi.org/10.1016/j.physletb.2016.11.047}
  {\bibfield  {journal} {\bibinfo  {journal} {Phys. Lett. B}\ }\textbf
  {\bibinfo {volume} {764}},\ \bibinfo {pages} {289} (\bibinfo {year}
  {2017})},\ \Eprint {https://arxiv.org/abs/1609.01253} {arXiv:1609.01253
  [gr-qc]} \BibitemShut {NoStop}%
\bibitem [{\citenamefont {Tukhashvili}\ and\ \citenamefont
  {Steinhardt}(2023)}]{Tukhashvili:2023itb}%
  \BibitemOpen
  \bibfield  {author} {\bibinfo {author} {\bibfnamefont {G.}~\bibnamefont
  {Tukhashvili}}\ and\ \bibinfo {author} {\bibfnamefont {P.~J.}\ \bibnamefont
  {Steinhardt}},\ }\href {https://doi.org/10.1103/PhysRevLett.131.091001}
  {\bibfield  {journal} {\bibinfo  {journal} {Phys. Rev. Lett.}\ }\textbf
  {\bibinfo {volume} {131}},\ \bibinfo {pages} {091001} (\bibinfo {year}
  {2023})},\ \Eprint {https://arxiv.org/abs/2307.16098} {arXiv:2307.16098
  [gr-qc]} \BibitemShut {NoStop}%
\bibitem [{\citenamefont {Chen}\ and\ \citenamefont
  {Wang}(2010)}]{Chen:2009zp}%
  \BibitemOpen
  \bibfield  {author} {\bibinfo {author} {\bibfnamefont {X.}~\bibnamefont
  {Chen}}\ and\ \bibinfo {author} {\bibfnamefont {Y.}~\bibnamefont {Wang}},\
  }\href {https://doi.org/10.1088/1475-7516/2010/04/027} {\bibfield  {journal}
  {\bibinfo  {journal} {JCAP}\ }\textbf {\bibinfo {volume} {04}},\ \bibinfo
  {pages} {027}},\ \Eprint {https://arxiv.org/abs/0911.3380} {arXiv:0911.3380
  [hep-th]} \BibitemShut {NoStop}%
\bibitem [{\citenamefont {Renaux-Petel}\ and\ \citenamefont
  {Turzy\'nski}(2016)}]{Renaux-Petel:2015mga}%
  \BibitemOpen
  \bibfield  {author} {\bibinfo {author} {\bibfnamefont {S.}~\bibnamefont
  {Renaux-Petel}}\ and\ \bibinfo {author} {\bibfnamefont {K.}~\bibnamefont
  {Turzy\'nski}},\ }\href {https://doi.org/10.1103/PhysRevLett.117.141301}
  {\bibfield  {journal} {\bibinfo  {journal} {Phys. Rev. Lett.}\ }\textbf
  {\bibinfo {volume} {117}},\ \bibinfo {pages} {141301} (\bibinfo {year}
  {2016})},\ \Eprint {https://arxiv.org/abs/1510.01281} {arXiv:1510.01281
  [astro-ph.CO]} \BibitemShut {NoStop}%
\bibitem [{\citenamefont {Senatore}\ \emph {et~al.}(2010)\citenamefont
  {Senatore}, \citenamefont {Smith},\ and\ \citenamefont
  {Zaldarriaga}}]{Senatore:2009gt}%
  \BibitemOpen
  \bibfield  {author} {\bibinfo {author} {\bibfnamefont {L.}~\bibnamefont
  {Senatore}}, \bibinfo {author} {\bibfnamefont {K.~M.}\ \bibnamefont
  {Smith}},\ and\ \bibinfo {author} {\bibfnamefont {M.}~\bibnamefont
  {Zaldarriaga}},\ }\href {https://doi.org/10.1088/1475-7516/2010/01/028}
  {\bibfield  {journal} {\bibinfo  {journal} {JCAP}\ }\textbf {\bibinfo
  {volume} {1001}},\ \bibinfo {pages} {028}},\ \Eprint
  {https://arxiv.org/abs/0905.3746} {arXiv:0905.3746 [astro-ph.CO]}
  \BibitemShut {NoStop}%
\bibitem [{\citenamefont {Cabass}\ \emph {et~al.}(2024)\citenamefont {Cabass},
  \citenamefont {Philcox}, \citenamefont {Ivanov}, \citenamefont {Akitsu},
  \citenamefont {Chen}, \citenamefont {Simonovi\'c},\ and\ \citenamefont
  {Zaldarriaga}}]{Cabass:2024wob}%
  \BibitemOpen
  \bibfield  {author} {\bibinfo {author} {\bibfnamefont {G.}~\bibnamefont
  {Cabass}}, \bibinfo {author} {\bibfnamefont {O.~H.~E.}\ \bibnamefont
  {Philcox}}, \bibinfo {author} {\bibfnamefont {M.~M.}\ \bibnamefont {Ivanov}},
  \bibinfo {author} {\bibfnamefont {K.}~\bibnamefont {Akitsu}}, \bibinfo
  {author} {\bibfnamefont {S.-F.}\ \bibnamefont {Chen}}, \bibinfo {author}
  {\bibfnamefont {M.}~\bibnamefont {Simonovi\'c}},\ and\ \bibinfo {author}
  {\bibfnamefont {M.}~\bibnamefont {Zaldarriaga}},\ }\href@noop {} {\
  (\bibinfo {year} {2024})},\ \Eprint {https://arxiv.org/abs/2404.01894}
  {arXiv:2404.01894 [astro-ph.CO]} \BibitemShut {NoStop}%
\bibitem [{\citenamefont {Sohn}\ \emph {et~al.}(2024)\citenamefont {Sohn},
  \citenamefont {Wang}, \citenamefont {Fergusson},\ and\ \citenamefont
  {Shellard}}]{Sohn:2024xzd}%
  \BibitemOpen
  \bibfield  {author} {\bibinfo {author} {\bibfnamefont {W.}~\bibnamefont
  {Sohn}}, \bibinfo {author} {\bibfnamefont {D.-G.}\ \bibnamefont {Wang}},
  \bibinfo {author} {\bibfnamefont {J.~R.}\ \bibnamefont {Fergusson}},\ and\
  \bibinfo {author} {\bibfnamefont {E.~P.~S.}\ \bibnamefont {Shellard}},\
  }\href {https://doi.org/10.1088/1475-7516/2024/09/016} {\bibfield  {journal}
  {\bibinfo  {journal} {JCAP}\ }\textbf {\bibinfo {volume} {09}},\ \bibinfo
  {pages} {016}},\ \Eprint {https://arxiv.org/abs/2404.07203} {arXiv:2404.07203
  [astro-ph.CO]} \BibitemShut {NoStop}%
\bibitem [{\citenamefont {Pimentel}\ and\ \citenamefont
  {Wang}(2022)}]{Pimentel:2022fsc}%
  \BibitemOpen
  \bibfield  {author} {\bibinfo {author} {\bibfnamefont {G.~L.}\ \bibnamefont
  {Pimentel}}\ and\ \bibinfo {author} {\bibfnamefont {D.-G.}\ \bibnamefont
  {Wang}},\ }\href@noop {} {\  (\bibinfo {year} {2022})},\ \Eprint
  {https://arxiv.org/abs/2205.00013} {arXiv:2205.00013 [hep-th]} \BibitemShut
  {NoStop}%
\bibitem [{\citenamefont {Jazayeri}\ \emph {et~al.}(2023)\citenamefont
  {Jazayeri}, \citenamefont {Renaux-Petel},\ and\ \citenamefont
  {Werth}}]{Jazayeri:2023xcj}%
  \BibitemOpen
  \bibfield  {author} {\bibinfo {author} {\bibfnamefont {S.}~\bibnamefont
  {Jazayeri}}, \bibinfo {author} {\bibfnamefont {S.}~\bibnamefont
  {Renaux-Petel}},\ and\ \bibinfo {author} {\bibfnamefont {D.}~\bibnamefont
  {Werth}},\ }\href {https://doi.org/10.1088/1475-7516/2023/12/035} {\bibfield
  {journal} {\bibinfo  {journal} {JCAP}\ }\textbf {\bibinfo {volume} {12}},\
  \bibinfo {pages} {035}},\ \Eprint {https://arxiv.org/abs/2307.01751}
  {arXiv:2307.01751 [hep-th]} \BibitemShut {NoStop}%
\bibitem [{\citenamefont {Babich}\ \emph {et~al.}(2004)\citenamefont {Babich},
  \citenamefont {Creminelli},\ and\ \citenamefont
  {Zaldarriaga}}]{Babich:2004gb}%
  \BibitemOpen
  \bibfield  {author} {\bibinfo {author} {\bibfnamefont {D.}~\bibnamefont
  {Babich}}, \bibinfo {author} {\bibfnamefont {P.}~\bibnamefont {Creminelli}},\
  and\ \bibinfo {author} {\bibfnamefont {M.}~\bibnamefont {Zaldarriaga}},\
  }\href {https://doi.org/10.1088/1475-7516/2004/08/009} {\bibfield  {journal}
  {\bibinfo  {journal} {JCAP}\ }\textbf {\bibinfo {volume} {0408}},\ \bibinfo
  {pages} {009}},\ \Eprint {https://arxiv.org/abs/astro-ph/0405356}
  {arXiv:astro-ph/0405356} \BibitemShut {NoStop}%
\bibitem [{\citenamefont {Biagetti}\ \emph {et~al.}(2017)\citenamefont
  {Biagetti}, \citenamefont {Dimastrogiovanni},\ and\ \citenamefont
  {Fasiello}}]{Biagetti:2017viz}%
  \BibitemOpen
  \bibfield  {author} {\bibinfo {author} {\bibfnamefont {M.}~\bibnamefont
  {Biagetti}}, \bibinfo {author} {\bibfnamefont {E.}~\bibnamefont
  {Dimastrogiovanni}},\ and\ \bibinfo {author} {\bibfnamefont {M.}~\bibnamefont
  {Fasiello}},\ }\href {https://doi.org/10.1088/1475-7516/2017/10/038}
  {\bibfield  {journal} {\bibinfo  {journal} {JCAP}\ }\textbf {\bibinfo
  {volume} {10}},\ \bibinfo {pages} {038}},\ \Eprint
  {https://arxiv.org/abs/1708.01587} {arXiv:1708.01587 [astro-ph.CO]}
  \BibitemShut {NoStop}%
\bibitem [{\citenamefont {Dimastrogiovanni}\ \emph {et~al.}(2018)\citenamefont
  {Dimastrogiovanni}, \citenamefont {Fasiello},\ and\ \citenamefont
  {Tasinato}}]{Dimastrogiovanni:2018uqy}%
  \BibitemOpen
  \bibfield  {author} {\bibinfo {author} {\bibfnamefont {E.}~\bibnamefont
  {Dimastrogiovanni}}, \bibinfo {author} {\bibfnamefont {M.}~\bibnamefont
  {Fasiello}},\ and\ \bibinfo {author} {\bibfnamefont {G.}~\bibnamefont
  {Tasinato}},\ }\href {https://doi.org/10.1088/1475-7516/2018/08/016}
  {\bibfield  {journal} {\bibinfo  {journal} {JCAP}\ }\textbf {\bibinfo
  {volume} {08}},\ \bibinfo {pages} {016}},\ \Eprint
  {https://arxiv.org/abs/1806.00850} {arXiv:1806.00850 [astro-ph.CO]}
  \BibitemShut {NoStop}%
\bibitem [{\citenamefont {Kolb}\ \emph {et~al.}(2023)\citenamefont {Kolb},
  \citenamefont {Ling}, \citenamefont {Long},\ and\ \citenamefont
  {Rosen}}]{Kolb:2023dzp}%
  \BibitemOpen
  \bibfield  {author} {\bibinfo {author} {\bibfnamefont {E.~W.}\ \bibnamefont
  {Kolb}}, \bibinfo {author} {\bibfnamefont {S.}~\bibnamefont {Ling}}, \bibinfo
  {author} {\bibfnamefont {A.~J.}\ \bibnamefont {Long}},\ and\ \bibinfo
  {author} {\bibfnamefont {R.~A.}\ \bibnamefont {Rosen}},\ }\href
  {https://doi.org/10.1007/JHEP05(2023)181} {\bibfield  {journal} {\bibinfo
  {journal} {JHEP}\ }\textbf {\bibinfo {volume} {05}},\ \bibinfo {pages}
  {181}},\ \Eprint {https://arxiv.org/abs/2302.04390} {arXiv:2302.04390
  [astro-ph.CO]} \BibitemShut {NoStop}%
\end{thebibliography}%
\end{document}